\documentclass[aps, twocolumn, nofootinbib,preprintnumbers, superscriptaddress]{revtex4-1}

\pdfoutput=1 

\usepackage{ragged2e}
\newcommand{\jcaption}[1]{\caption{\parbox{\linewidth}{\justifying #1}}}

\usepackage[font=small,justification=justified,singlelinecheck=false]{caption}
\usepackage{amsmath}      
\usepackage{mathtools}    
\usepackage{amssymb}
\usepackage{graphicx,tikz} 
\usetikzlibrary{calc}

\definecolor{CiteRed}{HTML}{C00000} 
\definecolor{EqRefBlue}{HTML}{1F4B99}
\definecolor{CiteGreen}{HTML}{0B7A0B}

\usepackage{tabularx}   
\usepackage{booktabs}   
\usepackage{array}      %
\newcolumntype{C}{>{\centering\arraybackslash}X}  
\newcolumntype{P}[1]{>{\centering\arraybackslash}p{#1}} 
\newcolumntype{M}[1]{>{\centering\arraybackslash}m{#1}}

\usepackage{tensor}
\usepackage{braket}
\usepackage{comment}
\usepackage{subcaption}
\usepackage{dsfont}
\usepackage{bbm}

\usepackage{upgreek}

\usepackage[normalem]{ulem}
\usepackage{amsmath}

\usepackage{enumerate}
\usepackage{epsfig}
\usepackage{mathbbol}

\def\Im{\mathop{\rm Im} }
\def\Re{\mathop{\rm Re} }

\newcommand\half{{\ensuremath{\frac{1}{2}}}}

\newcommand{\be}{\begin{equation}}
\newcommand{\ee}{\end{equation}}
\newcommand{\bea}{\begin{eqnarray}}
\newcommand{\eea}{\end{eqnarray}}
\newcommand{\bega}{\begin{gather}}
\newcommand{\eega}{\end{gather}}
\newcommand{\nn}{\nonumber\\}
\newcommand{\bi}{\begin{itemize}}
\newcommand{\ei}{\end{itemize}}
\newcommand{\ben}{\begin{enumerate}}
\newcommand{\een}{\end{enumerate}}
\newcommand{\bca}{\begin{cases}}
\newcommand{\eca}{\end{cases}}
\newcommand{\bln}{\begin{align}}
\newcommand{\eln}{\end{align}}
\newcommand{\bst}{\begin{split}}
\newcommand{\est}{\end{split}}
\def\ie{\begin{equation}\begin{aligned}}
\def\fe{\end{aligned}\end{equation}}
\newcommand{\bma}{\le(\begin{matrix}}
\newcommand{\ema}{\end{matrix}\ri)}
\newcommand{\bwt}{\begin{widetext}}
\newcommand{\ewt}{\end{widetext}}

\newcommand\ha{{\half}}

\def\le{\left}
\def\ri{\right}

\newcommand\sH{{\ensuremath{{\mathcal H}}}}

\newcommand{\rb}{{\rm{b}}}

\newcommand{\Tr}{\text{Tr}}

\usepackage{amsthm}
\usepackage{thmtools, thm-restate} 
\usepackage[justification=justified,singlelinecheck=false]{caption}

\newcommand{\Ai}{{\rm Ai}}

\newcommand{\lr}[1]{\left( #1 \right)}

\def\rb{\rangle}
\def\lb{\langle}

\def\qq{\qquad}

\usepackage[percent]{overpic}

\usepackage[colorlinks=true,
            linkcolor=EqRefBlue,   
            citecolor=CiteRed,     
            urlcolor=CiteGreen     
           ]{hyperref}

\begin{document}

\title{Negative energies and the breakdown of bulk geometry}

\author{John Preskill}
\affiliation{Institute for Quantum Information and Matter, California Institute of Technology,
Pasadena, CA 91125}

\author{Mykhaylo Usatyuk}
\email{musatyuk@kitp.ucsb.edu}
\affiliation{Kavli Institute for Theoretical Physics, Santa Barbara, CA 93106, USA}

\author{Shreya Vardhan}
\email{svardhan@caltech.edu}
\affiliation{Institute for Quantum Information and Matter, California Institute of Technology,
Pasadena, CA 91125}

\begin{abstract}

 \noindent 

 One central question in quantum gravity is to understand how and why predictions from semiclassical gravity can break down in regimes with low spacetime curvature. One diagnostic of such a breakdown is that states which are orthonormal at the semiclassical level can receive large corrections to their inner products from quantum fluctuations. We study this effect by examining inner products in pure 2D JT gravity. 
 Previous work showed that black hole states with long interiors exhibit a breakdown at length scales of order $e^{S_0}$, where $S_0$ is a parameter analogous to $1/G_N$ in higher dimensions. This breakdown is caused by the discreteness of the  spectrum of the dual boundary random matrix theory. We show that the full sum over quantum fluctuations 
 indicates a more dramatic breakdown at parametrically shorter lengths of order $e^{S_0/3}$. In the dual boundary description, these corrections arise from negative energy states appearing in rare members of the random matrix ensemble. 
These results demonstrate that non-perturbative effects can invalidate the semiclassical description at much smaller length scales than previously expected, providing a new mechanism for the breakdown of effective gravitational theories.

\end{abstract}


\maketitle

\section{Introduction}



According to conventional wisdom,  general relativity and quantum field theory provide a valid approximation to quantum gravity in regimes where spacetime curvatures are small compared to the Planck scale. In this approximation, the quantum states are specified by a spatial metric together with the profile of matter fields, $|\phi\rb \equiv |h_{i j}, \psi\rb$ \cite{DeWitt:1967yk}. Let us refer to this approximation as the {\it semiclassical} or {\it effective} description of gravity, and denote its Hilbert space by $\sH_{\rm eff}$. 

It is by now believed that there are many situations where this description must break down even with small spacetime curvatures. The most well-known example is Hawking's black hole information paradox \cite{hawking1, hawking2}, where the effective description is in conflict with the unitarity of quantum mechanics. Recent work \cite{penington, aemm, islands, pssy,  hartman_replica} has shown that the breakdown of the effective description can arise from new contributions to the gravitational path integral. However, the precise criteria and mechanism for the breakdown at the level of the Hilbert space remain to be fully understood.


In asymptotically AdS spacetimes, the AdS/CFT correspondence allows us to study the breakdown at the level of the Hilbert space using the ``bulk-to-boundary map''
\be  \label{Vmap}
V: \sH_{\rm eff} \to \sH_{{\rm bdry}, N}\, \cong  \sH_{{\rm bulk}, G_N = 1/N^2} \, .  
\ee
$V$ maps states in the effective description, which corresponds to the bulk gravity theory in the small $G_N$ limit, to states in the boundary CFT Hilbert space at finite $N$. $\sH_{{\rm bdry}, N}$ is believed to be isomorphic to a bulk quantum gravity Hilbert space $\sH_{{\rm bulk}, G_N}$ at finite $G_N$, although our understanding of the latter is much less explicit. We will  refer to both $\sH_{{\rm bdry}, N}$ and $\sH_{{\rm bulk}, G_N}$ interchangeably as the {\it fundamental} Hilbert space. Within this framework, the breakdown of the effective description for states $\ket{\phi},  \ket{\phi'}\in \sH_{\rm eff}$ is quantified by the difference 
\be \label{Vdiff}
\braket{\phi|V^{\dagger}V|\phi'}- \braket{\phi|\phi'}_{\text{eff}}\, . 
\ee
The quantity \eqref{Vdiff} can be seen as measuring the effects of all perturbative and non-perturbative quantum gravity fluctuations.\footnote{The first term in \eqref{Vdiff} is the Hilbert space interpretation of gravitational path integral prescriptions for quantum overlaps between bulk states  used for instance in~\cite{pssy, saad, nonpert, saad, firewalls, magan}.}
The key questions we should address in order to probe the breakdown of $\sH_{\rm eff}$ are: 
\begin{enumerate}
\item  How large can the difference in \eqref{Vdiff} 
 be? 
\item What are the geometric properties of  states $\ket{\phi}$ where the difference is large? 
\item From a microscopic Hilbert space perspective, when and why does \eqref{Vdiff} become large? 
\end{enumerate}

\begin{figure}[!t]
    \centering
    \includegraphics[width=\linewidth]{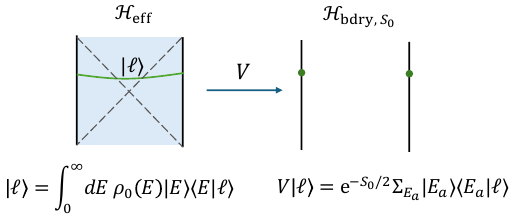}
    \caption{Bulk and boundary  descriptions of $\ket{\ell}$ states in JT gravity. }
    \label{fig:ell_def}
\end{figure}

In this work, we point out a new mechanism for the breakdown of the effective description of gravity from strong quantum fluctuations. We address questions 1-3 above in the theory of pure 2D JT gravity, taking  $\ket{\phi}$ to be the states $\ket{\ell}$ with fixed geodesic length $\ell$ between the two asymptotically AdS boundaries (see Fig.~\ref{fig:ell_def}). 
Our main result is that the inner products $\langle \ell | V^\dagger V | \ell' \rangle$ receive 
large corrections signaling a breakdown of semiclassical geometry when $\ell \sim e^{S_0/3}$, far smaller than the length $\ell \sim e^{S_0}$ which had been expected from earlier analyses ~\cite{volume_luca, nonpert, magan, akers, Miyaji:2024ity, Miyaji:2025yvm}. 

From the bulk perspective, we compute these corrections to the inner product by summing the gravitational path integral for JT gravity over all topologies. We also perform a computation from the boundary perspective that yields a matching result. The microscopic origin of these corrections from the boundary viewpoint is particularly striking. The dual description of JT gravity involves a random matrix ensemble, and we find that the dominant contribution to the breakdown comes from negative energy states that appear in rare members of the ensemble. Although such states occur with small probability, their contribution to inner products is exponentially enhanced at large $\ell$, and hence can become dominant (see Fig.~\ref{fig:spectrum}). This mechanism is distinct from previously studied effects arising from the discreteness of the energy spectrum in the boundary theory, and leads to a parametrically earlier breakdown of semiclassical physics. We will further show that negative energy states significantly modify classical expectations for the wormhole length in the eternal two-sided black hole geometry in JT gravity, leading to an infinite  expectation value of this length for all times.


\section{Setup and summary of results}

\noindent {\bf  Effective Hilbert space in JT gravity.} The  $\ket{\ell}$ states form an orthonormal basis for $\sH_{\rm eff}$ in pure JT gravity. The semiclassical inner product among these states is given by~\cite{harlow_jafferis, zhenbin}\footnote{In our notation, overlaps like \eqref{ell_orth} and \eqref{ell_e_mt} without insertions of $V$ are always between states in $\sH_{\rm eff}$.}
\be 
\braket{\ell|\ell'} = \delta(\ell-\ell') \, . \label{ell_orth} 
\ee   
An alternative  basis for $\sH_{\rm eff}$ consists of fixed energy states $\ket{E}$, which correspond to  black holes of energy $E$. The semiclassical inner product between $\ket{\ell}$ and $\ket{E}$ is
\be 
\braket{E|\ell} = 2\sqrt{2} K_{i \sqrt{8E}}(4 e^{-\ell/2})\, . \label{ell_e_mt} 
\ee
A parameter $S_0$ in JT gravity plays the role of $1/G_N$ in higher-dimensional theories. The above inner products \eqref{ell_orth}-\eqref{ell_e_mt} in $\sH_{\rm eff}$ include perturbative fluctuations  which can be expanded in powers of $G_N$, but do not include any nonperturbative effects of order $e^{-1/G_N}$.


\vspace{0.2cm}

\begin{figure}[!t]
\centering
\includegraphics[width=\columnwidth]{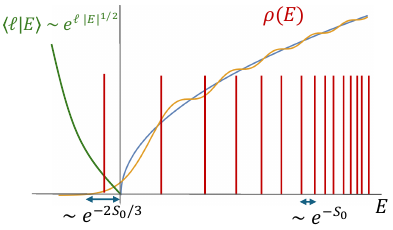}
\jcaption{\noindent Three versions of the density of states of JT gravity, $\rho(E) = e^{-S_0} \sum_{E_a} \delta(E-E_a)$. {\bf Red:} discrete spectrum for a given realization of the ensemble at finite $S_0$. {\bf Blue:} $S_0\to \infty$ limit of the ensemble average $\overline{\rho(E)}$, which matches the bulk density of states $\rho_0(E)$. {\bf Orange:}   $\overline{\rho(E)}$ at finite $S_0$ in the universal ``Airy'' regime. 
The key feature of the spectrum at finite $S_0$ underlying the breakdown in \eqref{deviation_result} and \eqref{variance_result} is the appearance of negative energies in some members of the ensemble, and ({\bf green}) the accompanying large value of $\braket{\ell|E}$ at negative energies.}
\label{fig:spectrum}
\end{figure}

\noindent {\bf Holographic map in JT gravity.} The boundary theory at finite $S_0$ should be thought of as a random Hamiltonian $H$ drawn from a random matrix ensemble~\cite{sss}.  A key difference between $\sH_{\rm eff}$ and $\sH_{{\rm bdry}, {S_0}}$ is that the former has a continuous spectrum  $|E\rb\in [0,\infty)$,\footnote{Note that this is analogous to the part of the spectrum above the black hole threshold in higher dimensions.} while the latter has a discrete spectrum $\{| E_a\rb  \}$ with typical level spacing $\sim e^{-S_0}$. The action of the map $V$ is given by sending the continuous bulk spectrum to the discrete boundary spectrum. Concretely, consider some state $\ket{\phi} \in \sH_{\rm eff}$, expressed in the bulk energy basis as $\ket{\phi}= \int_0^{\infty} dE \rho_0(E)\ket{E}\braket{E|\phi}$, where $\rho_0(E)$ is the bulk density of states.  The action of $V$ is given by~\cite{nonpert} (see Appendix~\ref{app:jt_review} for a review)
\be \label{Vdef_JT}
V \ket{\phi} =  e^{-S_0/2} \sum_{ E_a} \ket{ E_a}\braket{ E_a|\phi} \, .
\ee
In \eqref{Vdef_JT}, $\ket{E_a}$ denotes the energy eigenstate in the finite-dimensional boundary theory, and $\braket{ E_a|\phi}$ denotes the inner product in $\sH_{\rm eff}$.

Our goal in this paper is to address questions 1-3 of the introduction for the map $V$ in \eqref{Vdef_JT}, and in the bulk states $\ket{\ell}$. Given the randomness of the boundary Hamiltonian and consequently of $V$, it is also important to address a fourth question: 
\begin{enumerate}
    \item[4.] How large is the variance of $\braket{\ell|V^{\dagger}V|\ell'}$ between different members of the random matrix ensemble? 
\end{enumerate}

\noindent \textbf{Naive length scale for breakdown of overlaps.} Naively, it seems that the main source of a large value of the difference   \eqref{Vdiff} is that many states in $\sH_{\rm eff}$ get annihilated by $V$ in $\sH_{{\rm bdry}, S_0}$. These are known as ``null states" \cite{marolfmaxfield}. This is the main effect considered in existing literature~\cite{pssy, non_isometric, nonpert, magan, marolfmaxfield}. Ignoring the precise structure of $V$, one notices that with an energy cutoff, $\sH_{{\rm bdry}, S_0}$ has $\sim e^{S_0}$ states due to the discreteness of the spectrum, while $\sH_{\rm eff}$ is infinite-dimensional. In such cases, a simple model for $V$ consists of a random unitary matrix followed by a projection  \cite{non_isometric}. This model predicts that the difference \eqref{Vdiff} is small within any subspace of $\sH_{\rm eff}$ of  dimension $\lesssim e^{S_0}$, and in particular for any $\ell, \ell' \lesssim e^{S_0}$.\footnote{This statement assumes some discretization of the length in the presence of an energy cutoff.} This prediction appears to be consistent with calculations of related quantities in~\cite{volume_luca, magan}.

\vspace{0.2cm}

\noindent\textbf{New results for length breakdown.} On explicitly evaluating \eqref{Vdiff} and its variance for the $\ket{\ell}$ states in JT gravity, we find that both quantities start to become large at a much shorter length scale, when either length is $\ell \sim e^{S_0/3}$:\footnote{For all quantities in this paper, an overline indicates an average over the ensemble of Hamiltonians in JT gravity.}
\begin{align}
& \overline{\braket{\ell|V^{\dagger}V|\ell'}}- \braket{\ell|\ell'}  \sim  \exp \lr{\frac{4}{3}(\overline{\ell} e^{- S_0/3})^{3/2}}   \label{deviation_result}\\ 
& \overline{\braket{\ell|V^{\dagger}V|\ell'}^2} - \le(\overline{\braket{\ell|V^{\dagger}V|\ell'}}\ri)^2 \sim    \exp\lr{\frac{8\sqrt{2}}{3}(\overline{\ell} e^{- S_0/3})^{3/2}}  \label{variance_result}
\end{align}
where we defined the mean length 
$\bar \ell = \frac{1}{2}(\ell + \ell')$.\footnote{The $\sim$ notation indicates that we are not explicitly writing factors that do not scale exponentially with $\ell$ or $e^{S_0}$.}  We show in Sec.~\ref{sec:length_calculation} how the gravitational path integral expresses $\overline{\braket{\ell|V^{\dagger}V|\ell'}}$ and its variance as an infinite sum over 2d hyperbolic geometries of arbitrary genus. The sum over topologies  becomes tractable in the limit where $\bar\ell$ and $e^{S_0}$ are both large, while $\bar\ell e^{-S_0/3}\sim O(1)$. The exponential growth in $\bar \ell e^{-S_0/3}$ cannot be seen from a single geometry, and only arises from the full genus resummation. We derive \eqref{deviation_result} and \eqref{variance_result} in Sec.~\ref{sec:length_calcs} and Sec.~\ref{sec:variance_mt} respectively.

\vspace{0.2cm}

\noindent \textbf{Origin of breakdown from negative energies.}
From the perspective of the boundary theory, the origin of \eqref{deviation_result} and \eqref{variance_result} is unrelated to the discreteness of the spectrum, and is instead due to the appearance of negative energies in certain members of the matrix ensemble. Negative energies appear with small but $O(1)$ probability independent of $S_0$ among distinct draws of the ensemble, and have typical magnitude $\sim e^{-2S_0/3}$.  Crucially, for $E<0$,  the overlap $\braket{E|\ell}$ grows exponentially  $\sim e^{\ell \sqrt{|E|}}$. Hence, even a single component of the boundary state \eqref{Vdef_JT} containing a negative energy can significantly increase its norm for $\ell \sim e^{S_0/3}$. See Fig.~\ref{fig:spectrum}. The precise scaling in \eqref{deviation_result} and \eqref{variance_result} can be understood from a universal formula for the nonperturbative density of states close to a square root edge known in closed form in terms of the Airy function~\cite{TracyWidom1994Airy, eynard_review}, as we explain in Sec.~\ref{sec:airy_mt}.   The fact that the variance is significantly larger than the square of the correction to the average can be understood from the fact that in cases where negative energies do not appear in the spectrum, the value of $\braket{\ell|V^{\dagger}V|\ell'}$ is not exponentially large in $\ell^{3/2}$, and hence very far from the average value \eqref{deviation_result}. 

For larger lengths $\ell \sim e^{S_0}$ with $\ell e^{-S_0}\gg 1$, the relevant density of states no longer has the universal Airy form, and nonperturbative instanton effects specific to the JT gravity matrix integral become important. On including these effects, we find the behaviour 
\be
\overline{\braket{\ell|V^{\dagger}V|\ell'}} \sim e^{\bar\ell} \cos\le(\frac{\bar \ell\log  \bar\ell}{\pi}\ri), \quad \bar\ell e^{-S_0} \gg 1\, . \label{instanton}
\ee
The overall magnitude grows exponentially with $\bar\ell$, with large oscillations. Surprisingly, this quantity can be zero or negative even for $\ell = \ell'$, indicating that certain fine-tuned values of $\ell$ correspond to null states and negative norm states. We derive \eqref{instanton} in Sec.~\ref{sec:airy_mt} and Appendix~\ref{app:instanton}. 

\vspace{0.2cm}

\noindent\textbf{Expectation value of length in thermofield double.} We further discuss the dynamical implications of our results for the growth of the geodesic length between the two asymptotic boundaries in an eternal black hole geometry, which corresponds to the thermofield double (TFD) state $\ket{\psi_{\beta}} \equiv  \sum_{E} e^{-\beta E/2} \ket{E}$ in $\sH_{\rm eff}$. Classically, the two-sided length  grows linearly for all times. There has been extensive discussion in the literature of how this classical prediction is modified in the fundamental Hilbert space~\cite{complexity_action, volume_luca, nonpert, magan, akers, Miyaji:2024ity}.  
In Sec.~\ref{sec:tfd_mt}, we analyze the effect of negative energies on the expectation value in the fundamental description of the TFD state, $V\ket{\psi_{\beta}}$, of a definition $\hat \ell_{\rm fund}$ of the length operator in $\sH_{{\rm bdry}, S_0}$  previously used in~\cite{volume_luca}. This definition of $\hat \ell_{\rm fund}$ is natural from the perspective of the gravitational path integral. We will find that negative energies lead to a divergent expectation value of $\hat \ell_{\rm fund}$ for all times, 
\be 
\overline{\braket{\psi_{\beta+2it}|\hat \ell_{\rm fund} |\psi_{\beta+2it}}} = \infty \, .  \label{9}
\ee
The underlying mechanism for this divergence is an exponentially large contribution to \eqref{9} from lengths $\ell \gtrsim e^{S_0/3}$ for all $t$ (see \eqref{32}). 
Hence, a different definition of the fundamental length operator is necessary for the length expectation value in $V\ket{\psi_{\beta+2it}}$ to match classical predictions even at early times that are $O(1)$ in $e^{S_0}$.

\section{GENUS EXPANSION FOR LENGTH OVERLAPS} \label{sec:length_calculation}

In this section, we will derive an expression for $\overline{\braket{\ell|V^{\dagger}V|\ell'}}$ in JT gravity as a sum over topologies. From \eqref{Vdef_JT}, 
\be \label{V_rho}
\braket{\ell|V^{\dagger}V|\ell'}= \int dE \rho(E) \braket{\ell|E}\braket{E|\ell'}
\ee
where $\rho(E) = e^{-S_0} \sum_{E_a} \delta(E-E_a)$ is the normalized density of states for a single member of the ensemble at finite $S_0$. $\overline{\rho(E)}$  can be expressed in terms of the inverse Laplace transform of the averaged partition function $\overline{Z(\beta)}$, which can be equated with a sum over topologies through the gravitational path integral~\cite{sss}.  Combining this result with \eqref{V_rho}, we find (see Appendix~\ref{app:expansion_details} for details)  
\begin{align} 
\overline{\braket{\ell|V^{\dagger}V|\ell'}} &= \delta(\ell-\ell') +\sum_{g=1}^{\infty} e^{-2gS_0} \int_0^{\infty} db  \, b \,   K(\ell, \ell'; b) V_{g, 1}(b) \label{genus_sum_mt} \\
K(\ell, \ell'; b)&=2 K_0 \le(4\sqrt{e^{-\ell} + e^{-\ell'} + 2e^{-(\ell+\ell')/2} \cosh(b/2)}\ri) \nonumber\,.
\end{align}
where $V_{g, 1}(b)$ is a one-boundary Weil-Petersson volume~\cite{sss}, and $K_0$ is a modified Bessel function. The geometric interpretation of the parameter $b$ will be explained in \eqref{delta_diagrams} and point 3 below.  This sum is by definition equivalent to performing the Euclidean gravity path integral in JT gravity
\begin{align} \label{zgravdef_mt}
Z_{\rm JT} = \sum_{g=0}^\infty \int_{\text{ $\lb \ell|\ell'\rb$ b.c.}} \mathcal D g_{\mu \nu} \mathcal D \phi\, e^{-S_{\rm JT}[g_{\mu \nu}, \phi]}
\end{align}
where we integrate over the metric $g_{\mu \nu}$ and the dilaton $\phi$ with a particular set of boundary conditions and sum over the topological genus $g$ of the two-dimensional surface. The boundary conditions labelled by $\ell$,$\ell'$ correspond to geodesics that reach the asymptotic boundary of AdS, have renormalized lengths $\ell$ and $\ell'$, and meet at two points. The sum over $g$ in \eqref{genus_sum_mt} can be diagrammatically represented as follows, where in each diagram $b$ is the length of a closed geodesic along which the $\ell, \ell'$ boundary conditions are glued to the Weil-Petersson volume:
\begin{align} 
\label{delta_diagrams}
&\overline{\braket{\ell|V^{\dagger}V|\ell'}} - \delta(\ell-\ell') \nn &{\small =e^{-S_0}\int_0^{\infty} db \, b }\raisebox{-0.5cm}{\includegraphics[width=0.7\linewidth]{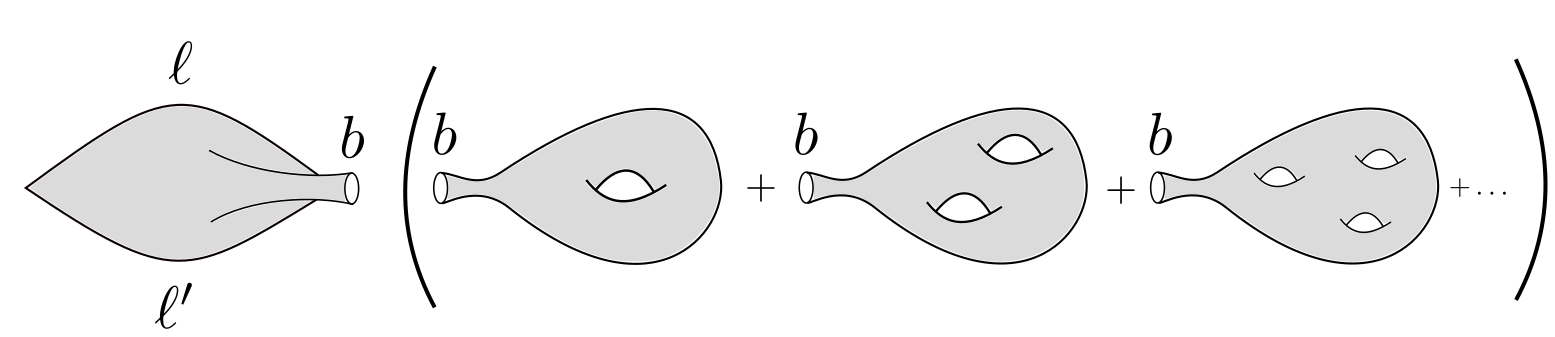}}
\end{align}

We now explain the various components of the sum in \eqref{genus_sum_mt} or \eqref{delta_diagrams}:  
\begin{enumerate}
\item The first term $\delta(\ell-\ell')$ in \eqref{genus_sum_mt}  matches the effective gravity answer~\eqref{ell_orth}. Its geometric origin is the $g=0$ disk topology, where the geodesic connecting two asymptotic boundary points is unique. Hence, geodesics with $\ell\neq \ell'$ cannot meet  at two endpoints on the boundary of the disk. 
\item On higher genus topologies, we can have two geodesics of any lengths $\ell$ and $\ell'$ connecting two asymptotic boundary points. Hence, the overlap between states with $\ell \neq \ell'$ becomes non-zero due to higher genus geometries. The probability that the spacetime fluctuates to a manifold of genus $g$ is exponentially suppressed as $e^{-2gS_0}$ due to a topological term in the action $S_{\rm JT}$. 
\item In the integral over $b$,  the Weil-Petersson volume $V_{g, 1}(b)$  ``counts'' how many topologically inequivalent hyperbolic two-dimensional manifolds there are of genus $g$ with one boundary of geodesic length $b$ \cite{sss, Mirzakhani:2006fta}.  The factor $K(\ell, \ell'; b)$ corresponds to the trumpet with three geodesic boundaries in \eqref{delta_diagrams}, and effectively restricts the range of contributing $b$ from $0$ to $\ell + \ell'$.   
 The leading  growth of  $V_{g,1}(b)$ for large $b$ is $\sim b^{6g-4}$. These two factors, together with the gluing measure $\int db \, b$, lead to a growth of the  integral proportional to $(\ell+ \ell')^{6g-1}$. 
\end{enumerate}

From points (2) and (3) above, the genus $g$ term in \eqref{genus_sum_mt} is proportional to $e^{-2gS_0} \le(\bar \ell\ri)^{6g-1}$, where  $\bar \ell \equiv \frac{\ell+\ell'}{2}$. When $\bar \ell$ becomes of order $e^{S_0/3}$, the large number of contributing manifolds of genus $g$ starts to compete with the small probability $e^{-2g S_0}$ that a higher genus surface fluctuates into existence. Hence, in this parameter regime, all orders in the sum over $g$ start to become important.

\section{Quantum corrections to  average of length overlaps}
\label{sec:length_calcs}

We now explicitly evaluate the quantum corrections to length inner products in \eqref{delta_diagrams}. For $\ell \ll e^{S_0/3}$, wormhole effects (i.e., contributions from higher-genus surfaces) are suppressed and unimportant, but at $\ell \sim e^{S_0/3}$ a subset of contributions become important and can be re-summed, as we explain below in Section~\ref{sec:bulk_length_calc}. In Sec~\ref{sec:airy_mt}, we match this bulk result for $\ell \sim e^{S_0/3}$ from the boundary perspective using the universal Airy density of states. We further predict a different behavior for large $\ell\sim e^{S_0}$ using non-perturbative instanton contributions.

\subsection{Sum over topologies in the Airy limit: $\ell \sim e^{S_0/3}$} \label{sec:bulk_length_calc}


Let us turn to explicit evaluation of the sum \eqref{genus_sum_mt}.  We show in Appendix~\ref{app:length_details} that evaluating the integral over $b$ gives
\begin{align} 
&\overline{\braket{\ell|V^{\dagger}V|\ell'}} - \lb \ell | \ell' \rb  = \sum_{g=1}^{\infty} L^{6 g} \sum_{p=0}^{3g-2}  (\bar\ell)^{-2p - 1} C_{g, p} , \nn
  &L \equiv \bar \ell e^{-S_0/3}\, \label{V_p_sum}
\end{align}
where $C_{g, p}$ is a constant independent of $\ell, e^{S_0}$. 
The sum over $p$ gives a factor $\sim g!$, so that \eqref{V_p_sum} is a divergent asymptotic series. However, each individual term is exponentially suppressed for $\bar\ell \ll e^{S_0/3}$, so that we do not expect large corrections in this regime. Crucially, for $\bar\ell \sim e^{S_0/3}$, the subset of $p=0$ terms become larger than all others and can be re-summed. This subset of terms is very similar to the one that survives in the well-known Airy limit.\footnote{The standard definition of the Airy limit is for the thermal partition function $\overline{Z(\beta)}$ where $\beta\sim e^{2 S_0/3}\to \infty$ while $\beta e^{-2 S_0/3}$ is fixed, see for instance  \cite{netta_free_energy, Saad:2022kfe, sergio, Hernandez-Cuenca:2024xlg, airy_tale}. In our case, an Airy-like limit arises from taking $e^{S_0} \to \infty, \quad \bar \ell \to \infty, \quad  \bar\ell e^{-S_0/3}   =  O(1) \,$. In this sense, $\ell^2$ plays a similar role to $\beta$.} Ignoring $p\geq1$ and performing the sum, we find: 
\begin{align}
\overline{\braket{\ell|V^{\dagger}V|\ell'}} &= \frac{1}{\bar\ell} \sum_{g=1}^{\infty} \frac{\le(\frac{L^6}{3}\ri)^{g} }{g! (3g-2)! (6g-2)(6g-1)} \, . 
\label{final_airysum} \\ 
& =\frac{L^6}{60\bar\ell} ~ {}_1F_4\le(\frac{5}{6};\frac{4}{3}, \frac{5}{3}, \frac{11}{6}, 2; \frac{L^6}{81} \ri)  \, . \label{1F4_growth}\\
&\sim \exp \lr{\frac{4}{3} \le(\bar\ell e^{-S_0/3}\ri)^{3/2}}, \quad \bar\ell e^{-S_0/3} \gg 1  \label{final_exp}
\end{align}
The importance of summing over all genus can be seen by comparing \eqref{final_exp} to the $g=1$ term in \eqref{V_p_sum},  
\be 
\overline{\braket{\ell|V^{\dagger}V|\ell'}}  \propto e^{-2S_0} (\bar\ell)^{5} \, .  \label{leading_correction}
\ee
Based on \eqref{leading_correction}, we would expect that the corrections become significant at $\bar \ell\sim e^{2S_0/5}$. In order to see the exponential growth with $\bar\ell$ for  $\bar \ell \sim e^{S_0/3}$, the full sum over genus is necessary. 

\textbf{Closed universes.} In most of the paper, we consider the sector of the JT gravity Hilbert space with two asymptotic boundaries. It is natural to ask whether the semiclassical inner products break down on a similar length scale if one considers the sector of the Hilbert space with no asymptotic boundaries, consisting of ``closed universe'' states. In pure JT gravity, these closed universe states are again labelled by their length $\ket{b}$, which can take values from 0 to $\infty$, and at the semiclassical level
they obey the orthonormality relation
$\braket{b|b'}= \frac{1}{b}\delta(b-b')$.  We analyze corrections to closed universe inner products from the sum over topologies in Appendix~\ref{app:closeduniversevar}, and find 
\be 
\lb b |V^{\dagger}V| b' \rb - \braket{b|b'} \sim \exp\lr{\frac{2\sqrt{2}}{3}(\bar b e^{-S_0/3})^{3/2}}
\label{19}\ee 
where $\bar b = \frac{1}{2}(b + b')$. Hence the semiclassical effective description of the bulk geometry breaks down for $\bar b \sim e^{S_0/3}$ in the case of a closed universe, much as it breaks down for $\bar \ell \sim e^{S_0/3}$ in the case with two asymptotic boundaries.
\footnote{On taking into account all quantum corrections for closed universes, one finds that there is a more drastic modification to the above inner product for all $b$ such that the Hilbert space is one-dimensional~\cite{islands,pssy,marolfmaxfield,McNamara:2020uza,Usatyuk:2024mzs}. This conclusion can be avoided by introducing an ``observer prescription''~\cite{Harlow:2025pvj,Abdalla:2025gzn}. Even with these observer prescriptions, the inner products should still receive the large corrections in \eqref{19} for $\bar b \sim e^{S_0/3}$.}


\subsection{Origin of large corrections from negative energies}
\label{sec:airy_mt}


{\bf Airy regime ($\ell \sim e^{S_0/3}$).} We now explain the origin of the exponential growth \eqref{final_exp} from the energy integral in \eqref{V_rho}. JT gravity is dual to a random matrix theory where the leading order density of states has square root edge behavior $\rho_0(E) \sim \sqrt{E}$ for small $E$. Matrix models with this leading-order edge behavior have a universal behaviour of the density of states to all orders in $e^{S_0}$, within a small energy window around $E=0$, in terms of the Airy function~\cite{eynard_review}.
To state this universal form, it is convenient to define a change of variables to $\lambda = 2^{1/3} E$. 
Then in terms of a variable  
$\xi \equiv -e^{2S_0/3}\lambda$, 
the density of states for $\lambda$ is
\be 
\rho_{\rm Airy}(\lambda) =  e^{-S_0/3}(\Ai'(\xi)^2 - \xi \Ai(\xi)^2 )  \label{airydef_mt} \, . 
\ee
The above expression holds at all orders in $e^{-S_0}$ as long as $\xi\sim O(1)$ implying $E\sim O(e^{-2 S_0/3})$, see Fig.~\ref{fig:spectrum}. Crucially, it has support for $E<0$. The appearance of negative energies is a nonperturbative effect that cannot be seen at any finite order in the genus expansion $O(e^{-2gS_0})$~\cite{sss}.  

The integral in \eqref{V_rho} thus has a contribution around $|E|\approx e^{-2 S_0/3}$ given by the Airy density of states 
\begin{align}
\overline{\braket{\ell| V^{\dagger}V|\ell'}} & \supset  \int_{-\epsilon}^{\epsilon} d\lambda \rho_{\rm Airy}(\lambda) \braket{\ell|2^{-\frac{1}{3}}\lambda}  \braket{2^{-\frac{1}{3}}\lambda
|\ell'}\,, \label{V_airy_mt}
\end{align}
where the integral should be truncated at $\epsilon = k e^{-2 S_0/3}$ with $k \gg 1$ to ensure that the approximation \eqref{airydef_mt} is applicable. Let us focus on the $\lambda < 0$ region and expand the integrand in the regime where the parameter $\xi\gg 1$, but not scaling with $e^{S_0}$. We have 
\begin{align}
&\rho_{\rm Airy}(\lambda) \sim  e^{-\frac{4}{3} \le( e^{2S_0/3} (-\lambda) \ri)^{3/2}} , \quad e^{2S_0/3} (-\lambda) \gg 1\, .\label{rho_largeE_mt}\\
&\braket{\ell|2^{-\frac{1}{3}}\lambda} \sim e^{ 2^{1/3}\ell \sqrt{(-\lambda)}}, \quad \lambda < 0, \quad \ell \gg 1 \, . \label{K_largel_mt}
\end{align}
The second line comes from \eqref{ell_e_mt} for large $\ell$.
Putting these asymptotic forms into \eqref{V_airy_mt} and changing variables to $z=-\lambda$, we find 
\be 
\overline{\braket{\ell| V^{\dagger}V|\ell'}}_{{\rm negative}~E} \sim \int_0^{\epsilon} dz \, e^{-\frac{4}{3}(e^{2S_0/3}z)^{3/2}} \, e^{2^{1/3} (\ell+\ell')\sqrt{z}} \, . \label{overlap_mt}
\ee
The saddle-point lies at 
\be 
z^{\ast} = 2^{-2/3} \bar\ell e^{-S_0} \, .  \label{zstar}
\ee
For $\bar\ell \sim e^{S_0/3}$, this corresponds to a value of $|\lambda|\sim e^{-2S_0/3}$, where the universal form \eqref{airydef_mt}  is justified for JT gravity. 
Evaluating the integrand in \eqref{overlap_mt} at the saddle-point value, we find that \eqref{overlap_mt} precisely matches  \eqref{final_exp}. One can check that the contributions to \eqref{V_airy_mt} from $\lambda>0$ are negligible compared to the contribution from $\lambda<0$. 

We have thus found that in terms of the spectrum, the origin of the exponential growth in \eqref{final_exp} is from the support of the Airy density on exponentially small negative energies.

Implicitly, in extending the integral \eqref{V_airy_mt} to negative energies, we have made a non-trivial assumption about the definition of $V$ in \eqref{Vdef_JT} in cases where $E_a<0$ appears in the boundary spectrum. We have assumed that in such cases, instead of setting the components along negative $\ket{E_a}$ to zero, we should analytically continue $\braket{\ell|E}$ to $E<0$ and include the contribution along $\ket{E_a}$. In principle, both choices for the definition of $V$ are compatible with the basic requirements for $V$ reviewed in Appendix~\ref{app:jt_review}. However, the above calculation shows that only the  definition with analytic continuation can match the result from the bulk sum over topologies in Sec.~\ref{sec:bulk_length_calc}. 

\textbf{Non-perturbative instantons $(\ell \sim e^{S_0})$.}
For $\ell \sim e^{S_0}$, the saddle-point \eqref{zstar} lies beyond the universal Airy regime, and therefore the above calculation using \eqref{rho_largeE_mt} is no longer valid.  The Airy density of states must now be replaced with the full density of states in the negative $E$ region including instanton effects~\cite{sss}. 
  We analyze these instanton corrections in Appendix~\ref{app:instanton}, and find that they lead to the behaviour~\eqref{instanton} for $\ell \gg e^{S_0}$.  We leave a detailed bulk understanding of these instanton contributions to future work. A bulk  interpretation involving D-branes for similar instanton contributions  to other quantities  has previously been proposed in \cite{sss,sergio}.


\section{Variance of length overlaps}
\label{sec:variance_mt}

We now consider the variance  of $\braket{\ell|V^{\dagger}V|\ell'}$. We can use either the definition of the map $V$ or a direct bulk path integral prescription to derive the following topological expansion for ${\rm Var}(\ell, \ell') = \overline{\braket{\ell|V^{\dagger}V|\ell'}^2}- \le(\overline{\braket{\ell|V^{\dagger}V|\ell'}}\ri)^2$:  
\begin{widetext}
\begin{equation}
{\rm Var}(\ell, \ell') =e^{-2S_0}\int_0^{\infty} db_1 b_1 \int_0^{\infty} db_2 b_2\raisebox{-.95cm}{\includegraphics[width=0.65\textwidth]{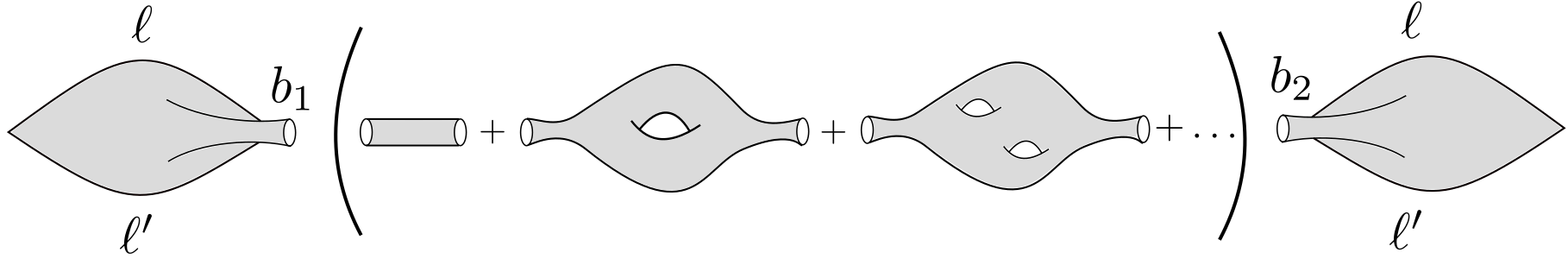}}
\label{var_sum}
\end{equation}
\end{widetext}
The first term in \eqref{var_sum} is analogous to the trumpet contribution to the spectral form factor~\cite{sss}. The remaining terms involve the Weil-Petersson volumes $V_{g, 2}(b_1, b_2)$ with two boundaries. Similar to the discussion below \eqref{V_p_sum}, for lengths scaling as $e^{S_0/3}$ we can keep a subset of contributions to the full sum which lead to a convrgent series.  
We analyze the sum in detail in Appendix~\ref{app:variance_details}. The result is 
\be \label{result_var}
{\rm Var}(\ell, \ell') =  \frac{1}{3}{\bar\ell}^4e^{-2S_0} \le[ 1 +  \sum_{g=1}^{\infty} \frac{1}{g!(3g)!} \le(\frac{64 L^6}{3}\ri)^g f(g) \ri]
\ee
where $f$ is some  function with a power law dependence on $g$. We further argue in Appendix~\ref{app:variance_details} that irrespective of the specific functional form of $f$, the above sum scales for large $L$ as  
\be 
{\rm Var}(\ell, \ell') \sim \exp\le(\frac{8\sqrt{2}}{3} \le(\bar\ell e^{-S_0/3}\ri)^{3/2}\ri) \, .  \label{var_ell}
\ee

It is useful to understand how this variance arises from properties of the  density of states. We have 
\begin{align}
&{\rm Var}(\ell, \ell')
= \nn 
&\int dE_1 dE_2 \overline{\rho(E_1)\rho(E_2) }_{\rm c} \langle\ell|E_1\rangle \langle E_1|\ell'\rangle\langle\ell|E_2\rangle\langle E_2|\ell'\rangle\label{cov_energy}
\end{align}
where $\overline{\rho(E_1)\rho(E_2) }_{\rm c}$  is the averaged connected two-point function of the density of states. Close to the square-root edge, the non-perturbative expression  for this connected two-point function again has a universal form, expressed in terms of the ``Airy kernel.''  Again, this two-point function is universal in a regime where the variable $\xi = e^{2S_0/3} E$ is $O(1)$, and the integrand in \eqref{cov_energy} 
has support on negative energies. On expanding the Airy kernel for negative $\xi$ with $|\xi|\gg 1$ and the factors of $\braket{\ell|E}$ for $\ell \gg 1$, we precisely reproduce the result \eqref{var_ell}. Again, the saddle-point values of the energies are negative and within the universal Airy regime. See Appendix~\ref{sec:variance_airy} for details of this calculation. 

Note that the first term in \eqref{result_var} starts to become large for $\bar\ell \sim e^{S_0/2}$. Since this contribution is analogous to the trumpet contribution to the spectral form factor, we expect that it should capture the effects of level repulsion in the spectrum. Indeed, the scaling $\ell \sim e^{S_0/2}$ for the breakdown of inner products is natural from the perspective of discreteness of the spectrum with spacings of $O(e^{-S_0})$, as $\braket{E|\ell}$ has an oscillatory form as a function of $E$ for $E>0$, with the period of oscillations $\sim 1/\ell^2$. However, this contribution to the variance from discreteness turns out to be negligible compared to the effect of possible negative energies captured by the full sum over $g$. 

We can see that, due to the larger prefactor inside the exponential, the variance in \eqref{var_ell} is much larger than the correction to the mean in \eqref{final_exp}. Physically, this large variance comes from the fact that whether or not negative energies arise in a particular member of the ensemble makes a dramatic difference to the value of $\braket{\ell|V^{\dagger}V|\ell'}$.

\section{Consequences for length growth in the TFD state}
\label{sec:tfd_mt}

We now discuss the dynamical consequences of our results. A long-standing dynamical question in quantum gravity is about the length of the wormhole in the two-sided asymptotically AdS eternal black hole, which corresponds to the thermofield double state
$\ket{\psi_{\beta}} = \sum_E e^{-\beta E/2} \ket{E}\in \sH_{\rm eff}$ \cite{maladacena_eternal}. In classical GR, the length of the spacelike geodesic connecting points at time $t$ on the two boundaries grows linearly for all time, $\ell_{\rm classical}(t) = \frac{2\pi}{\beta} t$. Based on the conjecture that bulk volume is dual to computational complexity in the boundary theory in AdS/CFT~\cite{Susskind:2014moa,Susskind:2014yaa, Stanford:2014jda}, it was proposed that there should be a mechanism for saturation of this length at $\ell, t \sim e^{S_0}$ when quantum corrections are included. In JT gravity, the disk-level semiclassical  calculation~\cite{zhenbin} reproduced the  classical result,
\be \label{ell_sc}
\braket{\ell(t)}_{\rm eff}  \equiv \int_{-\infty}^{\infty} d \ell\, \ell \,   |\braket{\psi_{\beta + 2it}|\ell}|^2 = \frac{2\pi}{\beta} t \, . 
\ee
In order to include certain effects from higher-genus topologies, \cite{volume_luca} considered the following quantity:\footnote{The definition in \cite{volume_luca} was introduced in a slightly different way using two-point functions of boundary operators, but precisely equals \eqref{luca_length}-\eqref{e1e2}.}   
\begin{align} \label{luca_length}
&\braket{\ell(t)}  = \int_{-\infty}^\infty d \ell\, \ell \,   \overline{|\braket{\psi_{\beta + 2it}|V^{\dagger}V|\ell}|^2} \, ,  \\  
&  \overline{|\braket{\psi_{\beta + 2it}|V^{\dagger}V|\ell}|^2}_{\rm positive~E} =\int_0^\infty  d E_1 d E_2  \nn 
& e^{-\frac{\beta}{2}(E_1 + E_2) - i t (E_1-E_2)}  
 \times \braket{E_1|\ell}\braket{\ell|E_2}\overline{\rho(E_1)  \rho(E_2)} \,. \label{e1e2} 
\end{align}
It is standard to approximate $\overline{\rho(E_1)  \rho(E_2)}$ by the sine-kernel ansatz, which is valid for $E_1, E_2 \gg 0$ while $|E_1-E_2| \ll 1$ \cite{mehta2004random}. Crucially, without including negative energies, $\braket{\ell(t)}$ saturates at $\ell, t \sim e^{S_0}$~\cite{volume_luca}. 



Our discussion so far shows that  it is important to extend the integrals in \eqref{e1e2} to negative energies.
For $\ell \sim e^{S_0/3}$, we can use the universal Airy form for  $\overline{\rho(E_1) \rho(E_2)}$, and like in the previous sections a saddle-point at negative energy leads to exponential growth of the inner product \eqref{e1e2} (see Appendix~\ref{app:tfd_airy})
\be  \label{32}
\overline{|\braket{\psi_{\beta + 2it}|V^{\dagger}V|\ell}|^2} \sim \exp\le({ \frac{2\sqrt{2}}{3} (e^{-S_0/3}\ell)^{3/2}}\ri) f(\ell, e^{S_0}, t)
\ee
where $f(\ell, e^{S_0}, t)$ is a prefactor that does not scale exponentially with powers of $\ell$ or $e^{S_0}$. Hence, $\ell \sim e^{S_0/3}$ and $E_1, E_2 <0$ give a dominant and very large contribution to the integral \eqref{luca_length} for any $t, \beta$, including when $t, \beta \sim O(1)$. For $\ell \gg e^{S_0/3}$, the Airy kernel is no longer valid, and the form of $\overline{\rho(E_1)\rho(E_2)}$ for larger energies in JT gravity is not known. Based on the instanton calculations in Appendix~\ref{app:instanton}, we expect that 
the inner product \eqref{e1e2}  continues to grow exponentially with $\ell$ in this regime, leading to a divergence in $\braket{\ell(t)}$ on integrating over all $\ell$.
There is no obvious way of subtracting the above divergence to obtain a meaningful finite time-dependent result which agrees with the classical result \eqref{ell_sc} even at early times.

Note that the above divergence  can be avoided with the length remaining finite and plateauing at $\ell \sim e^{S_0}$~\cite{volume_luca} when either (i) one considers a microcanonical version of the TFD state as in for instance~\cite{firewalls}, which by definition avoids any overlap with negative energy states, or (ii) one considers a version of JT gravity with a nonperturbative completion that does not have negative energies~\cite{johnson} (see the Discussion for further comments).

\section{Discussion}


In this work, we have shown that overlaps among states of fixed 
geodesic length $\ell$ in pure JT gravity receive large corrections 
from quantum fluctuations at parametrically shorter scales than 
previously expected. In particular, we find that these corrections 
become exponentially large when $\ell \sim e^{S_0/3}$, rather than 
at $\ell \sim e^{S_0}$ as suggested by previous arguments based on the finite 
dimensionality of the boundary Hilbert space or the saturation of boundary complexity.

The origin of this breakdown is a nonperturbative effect arising from 
the appearance of negative energy states in the boundary random matrix 
ensemble. Although such states occur only in rare realizations, their 
contribution to overlaps is exponentially enhanced at large $\ell$, 
allowing them to dominate expectation values. This mechanism is distinct 
from previously studied effects associated with the discreteness of the 
spectrum. 

From the bulk perspective, these effects correspond to a resummation 
over all topologies in the gravitational path integral, which becomes 
essential in the regime $\ell \sim e^{S_0/3}$. Our results therefore 
demonstrate that the breakdown of the semiclassical Hilbert space can 
be driven by nonperturbative contributions that are invisible in the 
perturbative expansion but dominate physical observables.

There are a number of directions for future exploration based on these results, which we summarize below. 

\textbf{Nonperturbative completions of JT gravity without negative energies.}   \cite{johnson} proposed an alternative random matrix model  to~\cite{sss}, which can also be seen as a  nonperturbative completion of the topological  expansion of JT gravity. In the model of~\cite{johnson}, only positive energies appear in the spectrum even at finite $S_0$. Hence, from the perspective of the boundary density of states, this model does not show the exponential growth of the length basis  overlaps in \eqref{deviation_result}-\eqref{variance_result}. It would be interesting to better understand the bulk reason for the lack of exponential growth in the model of~\cite{johnson}.\footnote{We thank Clifford Johnson for discussions on this point.} 

\textbf{Instanton corrections to bulk inner products.}
We see from \eqref{instanton} that at $\ell\gtrsim e^{S_0}$, the inner product \eqref{instanton} exponentially grows in overall magnitude, while also oscillating in sign and occasionally vanishing, giving rise to null states. The underlying mechanism behind these counterintuitive behaviors is the inclusion of complex energies in the the matrix model. It would be interesting to better understand the interpretation of these complex energies from the bulk perspective.



\textbf{Definition of the boundary length operator.}
The definition of $\braket{\ell(t)}$ in \eqref{luca_length} corresponds to evaluating the expectation value in $V\ket{\psi_{\beta+2it}}$ of the  operator  
$\hat \ell_{{\rm fund}} \equiv \int_{-\infty}^{\infty} d\ell \,  \ell  \,  V \ket{\ell}\bra{\ell}V^{\dagger}$. 
While $\hat \ell_{\rm fund}$ appears to be the natural definition of the fundamental length operator from the bulk path integral perspective, it has previously been pointed out~\cite{magan, akers, Miyaji:2024ity} that since the $V\ket{\ell}$ states do not form an orthonormal basis for $\sH_{{\rm bdry}, S_0}$, $\hat \ell_{{\rm fund}}$ is not a good definition from the Hilbert space perspective. A better definition  from the Hilbert space perspective would involve replacing the $V\ket{\ell}$ states with an orthonormal basis obtained from Gram-Schmidt orthonormalization, as previously suggested for instance in~\cite{Miyaji:2024ity}. However, the bulk path integral representation of this modified definition is likely to be complicated. Note that the existing  modified definitions of the $\hat \ell_{\rm fund}$ in~\cite{magan, akers, Miyaji:2024ity} do not take into account the effect of negative energies. 
We leave a more detailed discussion of a better definition of the length operator using Gram-Schmidt orthonormalization to future work.

\textbf{The role of complexity.}
Previously, the microscopic mechanism for the breakdown of bulk lengths for $\ell \sim e^{S_0}$ was understood in terms the  saturation of the  boundary complexity at late times~\cite{complexity_action, volume_luca, magan}. The mechanism based on negative energies found in this paper appears to be distinct from the mechanism based on complexity. It would be interesting to better understand the interplay between these different mechanisms for the breakdown of gravitational effective field theory. 

In the setup of evaporating black holes,~\cite{non_isometric} proposed that {\it bulk complexity} may be responsible for the breakdown of overlaps between semiclassical states in the presence of matter in the black hole interior. In order to test this proposal, it is important to extend the analysis of the present paper to the case of JT gravity with matter. One challenge in the case of JT with matter is to  regulate the  divergences appearing in higher-genus contributions to bulk overlaps~\cite{eth_jt}. 

\textbf{Higher dimensions.}~
It would be interesting to see whether the breakdown of the length basis found here for 2D JT gravity has an analog in higher dimensions. The most immediate candidate for a higher-dimensional setup which may show similar phenomena is the universal Schwarzian sector of 3D gravity~\cite{maxfield_turiaci1}, where it has been conjectured that an analogous sum over 3-geometries may reproduce the Airy density of states.


More broadly, our results suggest that the validity of semiclassical 
gravity is more fragile than previously appreciated. Rare 
nonperturbative spectral fluctuations can dominate physical observables 
and invalidate the effective description at parametrically earlier 
scales than those predicted by simple Hilbert space counting arguments. 
Understanding the role of such effects may be essential for a complete 
formulation of quantum gravity beyond the semiclassical approximation.

\vspace{0.2in} \centerline{\bf{Acknowledgments}} \vspace{0.2in}

We thank Temple He, Clifford Johnson, Luca Iliesiu, Thomas Schuster, David Simmons-Duffin, and Ronak Soni for helpful discussions. JP and SV acknowledge funding provided by the DOE QuantISED program
(DE SC0018407) and the Institute for Quantum Information and Matter, an NSF Physics
Frontiers Center (NSF Grant PHY-2317110). MU was supported in part by grant NSF PHY-2309135 to the Kavli Institute for Theoretical Physics (KITP), and by grants from the Simons Foundation (Grant Number 994312, DG), (216179, LB). This research was supported in part by grant NSF PHY-2309135 to the Kavli Institute for Theoretical Physics (KITP).

\onecolumngrid
\appendix

\section{Review of bulk-to-boundary map in JT gravity}
 \label{app:jt_review}

\subsection{Density of states and inner products in $\sH_{\rm eff}$}
In JT gravity, the effective Hilbert space $\sH_{\rm eff}$ is obtained from canonical quantization of the classical phase space of solutions to the equations of motion, or alternatively by ``cutting open'' the path integral  on the disk topology ~\cite{harlow_jafferis, zhenbin}. One complete orthonormal basis for $\sH_{\rm eff}$ consists of the eigenstates $\ket{E}$ of the ADM Hamiltonian $\hat H_{\rm  ADM}$, which generates time translations along the two asymptotic boundaries. The corresponding eigenvalues $E$ take all values in the continuum from 0 to $\infty$. $\hat H_{\rm ADM}$ does not commute with the length operator $\hat \ell$, whose eigenstates $\ket{\ell}$ have fixed geodesic length $\ell$ between the two boundaries. $\ell$ takes all values in the continuum from $-\infty$ to $\infty$.\footnote{Note that $\ell$ is a renormalized length in which we subtract the divergent contributions from near the AdS boundaries, and can take values from $-\infty$ to $\infty$ due to this subtraction. In practice, only lengths $>0$ are relevant for our discussion, as the overlaps $\ket{\ell|E}$ in \eqref{ell_e} are strongly suppressed for large negative $\ell$ for all $E$. }   The relation between the length and energy bases in $\sH_{\rm eff}$ is given by 
\be  \label{ell_e}
\braket{\ell|E} = 2 \sqrt{2} K_{i\sqrt{8E}}(4 e^{-\ell/2})\, .  
\ee
 The normalization in \eqref{ell_e} is chosen to be consistent with the following conventions for the inner products within the $\ket{\ell}$ or $\ket{E}$ bases:    
\be 
\braket{\ell|\ell'} = \delta(\ell- \ell'), \quad \braket{E|E'} = \frac{\delta(E-E')}{\rho_0(E)}\,  \label{norms} 
\ee
where 
\be \label{rho0_def} 
\rho_0(E) = \frac{1}{2\pi^2}\sinh(2\pi \sqrt{2E}) \, . 
\ee
We introduce $\rho_0(E)$
 above as it is the inverse Laplace transform of the disk partition function corresponding to $\Tr[e^{-\beta \hat H_{\rm ADM}}]$, and can therefore be interpreted as a density of states in $\sH_{\rm eff}$.

 From \eqref{norms}, the resolution of the identity for $\sH_{\rm eff}$ in the $\ket{E}$ basis is 
\be\mathbf{1}=\int dE \rho_0(E)\ket{E}\bra{E},
\ee
This is consistent with \eqref{ell_e} as it ensures  
\be 
\braket{\ell|\ell'}= \int_0^{\infty} dE \rho_0(E) \braket{\ell|E} \braket{E|\ell'} = \delta(\ell- \ell') \, . \label{ell_orth_app} 
\ee

From the Euclidean path integral perspective, 
there is a precise sense in which the inner product \eqref{ell_e} (and hence \eqref{ell_orth_app}) takes into account perturbative quantum fluctuations of the metric around the classical ${\rm AdS}_2$ solution at all orders in $1/S_0$ on the disk topology, as explained in~\cite{zhenbin, stanford_witten}. 

\subsection{Bulk-to-boundary map}

 In a single instance of the boundary random matrix ensemble at finite $S_0$, the Hilbert space is spanned by the discrete energy eigenstates  $\ket{ E_a}$ of $H$. We refer to this boundary Hilbert space at some given finite value of $S_0$ as $\sH_{{\rm bdry}, S_0}$. The states $\ket{E_a}$  form an orthonormal basis for $\sH_{{\rm bdry}, S_0}$, which satisfies   
$\braket{ E_a| E_b} = \delta_{ab}$. Note that the average of the normalized density of states for these energies approaches the semiclassical value \eqref{rho0_def} in the $S_0 \to \infty$ limit:
\be 
\lim_{S_0\to \infty}\overline{\rho(E)}  = \lim_{S_0\to \infty}\overline{e^{-S_0}\sum_{E_a}\delta(E-E_a)}  = \rho_0(E) \, .  \label{leading_density}
\ee

 Ref.~\cite{nonpert} made the  concrete proposal for the form of the bulk-to-boundary map $V$ in JT gravity given in \eqref{Vdef_JT}. 
Ref.~\cite{akers} proposed a simple and useful way of justifying this  choice of $V$, which we now review. The definition  \eqref{Vdef_JT} for  $V$ is almost uniquely determined by the following three requirements for $V$: 
\begin{enumerate}
\item $V$ is linear. This is a natural assumption that ensures consistency between the linearity of quantum-mechanical observables in $\sH_{\rm eff}$ and $\sH_{{\rm bdry}, S_0}$. 
\item Recall that $\sH_{\rm eff}$ and $\sH_{{\rm bdry}, S_0}$ both have their own versions of the dynamics along the boundary, generated respectively by $\hat H_{\rm ADM}$ and $H$. $V$ must be  consistent or ``equivariant'' with the dynamics of both $\sH_{\rm eff}$ and $\sH_{\rm bdry, S_0}$.  In particular, for some state $\ket{\psi}\in \sH_{\rm eff}$, the state that we obtain by first evolving $\ket{\psi}$ by $e^{-i\hat H_{\rm ADM}t}$ and then mapping to $\sH_{\rm bdry, S_0}$ using $V$ should be equal to the state $e^{-iHt}V\ket{\psi}$  from first sending $\ket{\psi}$ to $\sH_{{\rm bdry}, S_0}$ and then evolving by $H$. This corresponds to the condition 
\be 
V \hat H_{\rm ADM} = H V  \label{equivariance}
\ee
In order to meet this condition, we must have 
\be 
V\ket{E}  = \sum_{E_a} c_{ E_a} \ket{ E_a} \delta(E- E_a) \label{Vgen}
\ee
for some coefficients $c_{E_a}$. 
\item On averaging over the ensemble and taking the limit $e^{S_0}\to \infty$, the  overlaps in $\sH_{{\rm bdry}, S_0}$ should match those in $\sH_{\rm eff}$ from \eqref{norms}: 
\be \label{limit}
\text{lim}_{e^{S_0}\to \infty} \overline{\braket{E|V^{\dagger}V|E'}} = \braket{E|E'} = \frac{\delta(E-E')}{\rho_0(E)} \, . 
\ee
We can check using \eqref{leading_density} that this requirement for $V$ of the form \eqref{Vgen} fixes  $|c_{ E_a}|^2 = \frac{e^{-S_0}}{\rho_0( E_a)^2}$. Taking $c_{ E_a}$ to be real, we obtain 
\be 
V\ket{E}  = e^{-S_0/2}\sum_{E_a} \frac{1}{\rho_0(E)} \ket{ E_a} \delta(E- E_a) \label{Ve}
\ee
Combined with our first assumption of linearity, \eqref{Ve} precisely leads to \eqref{Vdef_JT} for a general state in $\sH_{\rm eff}$. 
\end{enumerate}

Note that one aspect of the definition of $V$ that is not fixed by the conditions 1-3 above is precisely how \eqref{Vdef_JT} should be interpreted in cases where some of the energies $E_a$ appearing in the finite $S_0$ boundary spectrum are negative, and the overlaps $\braket{E|\phi}$ have some continuation to negative $E$ which is non-zero. There are two choices: (i) we can include the components along $\ket{E_a}$ for $E_a<0$ and use the analytic continuation of $\braket{E|\phi}$ to $E<0$, or (ii) we can set the components along $E_a<0$ to zero. Both options (i) and (ii) are consistent with the conditions (i)-(iii) above.  It is the matching between our bulk and boundary calculations in Sec.~\ref{sec:length_calcs}  eventually justifies the choice (i), which we assume in this paper.

\section{Details of the derivation of the topological expansion for $\overline{\braket{\ell|V^{\dagger}V|\ell'}}$}
\label{app:expansion_details}

In this Appendix, we derive the topological expansion \eqref{genus_sum_mt} starting from the definition of $V$ in \eqref{Vdef_JT}, which implies that 
\be 
\label{ell_appendix}
\overline{\braket{\ell|V^{\dagger}V|\ell'}} = \int dE \, \overline{\rho(E)}\,  \braket{\ell|E}\braket{E|\ell'}
\ee
$\overline{\rho(E)}$ is proportional to the inverse Laplace transform of the averaged boundary partition function, $\overline{Z(\beta)} = \overline{\Tr[e^{-\beta H}]}$:  
\be 
\overline{\rho(E)} = e^{-S_0}\int_{-i \infty+ \gamma}^{i\infty+ \gamma} d\beta \, e^{\beta E} \,  \overline{Z(\beta)}\label{full_rhoE}
\ee
where $\gamma$ is chosen such that the integration contour is to the right of all singularities in $\overline{Z(\beta)}$.
From~\cite{sss}, $\overline{Z(\beta)}$ has the following expansion in powers of $e^{-S_0}$: 
\be 
\overline{Z(\beta)}_{\rm all~ orders} = Z_{\rm disk}(\beta) + \int_0^{\infty} db b Z_{\rm trumpet}(\beta, b)\sum_{g=1}^{\infty} V_{g, 1}(b)   e^{(1-2g)S_0} \label{zbeta_all}
\ee
where 
\be
Z_{\rm disk}(\beta)=e^{S_0}\frac{1}{(2\pi)^{\ha}(\beta)^{3/2}}  e^{\frac{2\pi^2 }{\beta}} , \quad Z_{\rm trumpet}(\beta, b) = \frac{e^{-b^2/(2\beta)}}{\sqrt{2\pi\beta}} \, , 
\ee
and $V_{1,g}$ are the Weil-Petersson volumes of genus $g$ with one boundary.   Putting this expansion into \eqref{full_rhoE}, we get 
\begin{align}
&\overline{\rho(E)} = \rho_0(E) + \sum_{g=1}^{\infty}  e^{(-2g)S_0} \rho_g(E), \label{expansion} \\ & \rho_g(E) \equiv \int_{-i\infty}^{i\infty} d\beta e^{\beta E}\int_0^{\infty} db\,  b   \frac{e^{-b^2/(2\beta)}}{\sqrt{2\pi \beta}} V_{g, 1}(b)\, .  \label{all_orders}
\end{align}
To obtain the first term of \eqref{expansion},  we have used the fact that the inverse Laplace transform of $e^{-S_0}Z_{\rm disk}(\beta)$ gives $\rho_0(E)$. Putting \eqref{all_orders} into \eqref{ell_appendix}, interchanging the order of the $b$, $E$ and $\beta$ integrals, and taking $\gamma \to 0$, we get:
\begin{align} 
\overline{\braket{\ell|V^{\dagger}V|\ell'}} - \delta(\ell-\ell') 
&~= \sum_{g=1}^{\infty} e^{-2gS_0} \int_0^{\infty} db\,  b\,  V_{g, 1}(b) \int_0^{\infty}dE \braket{\ell|E}\braket{E|\ell'} \int_{-i \infty}^{i\infty}  d\beta \frac{e^{\beta E -b^2/(2\beta)}}{\sqrt{2\pi \beta}}   \\ 
&~= \sum_{g=1}^{\infty} e^{-2gS_0} \int_0^{\infty} db\,  b\,  V_{g, 1}(b) \int_0^{\infty}dE \braket{\ell|E}\braket{E|\ell'}   \frac{\cos(\sqrt{2E} b)}{\sqrt{2E}\pi } \label{genus_sum_e}\\
&~=\sum_{g=1}^{\infty} e^{-2gS_0} \int_0^{\infty} db \, b \, V_{g, 1}(b) 2 K_0 \le(4\sqrt{e^{-\ell} + e^{-\ell'} + 2e^{-(\ell+\ell')/2} \cosh(b/2)}\ri)  \label{genus_sum}
\end{align}
where in the last step we have made use of a Bessel function identity. Note that in this last step, it is important that the integral  over $E$ goes from 0 to $\infty$. This is justified by the fact that each $\rho_g(E)$ has support only for $E>0$.\footnote{We will see in the next section that $V_{g, 1}(b)$ involves even powers of $b$. As a result, after evaluating the integral over $b$, each term in $\rho_g(E)$ has the form $\int_{-i\infty}^{i\infty} d\beta e^{\beta E} \beta^{n/2} $ for some odd integer $n>0$. These inverse Laplace transforms are defined only through analytic continuation, and are proportional to $E^{-1-n/2}$, which has a singularity at $E=0$. This subtlety with the definition of $\rho_g(E)$ ends up not being important, as the final integral over $b$ in \eqref{genus_sum} turns out to be finite and well-defined. Moreover, it precisely agrees with the gravitational path integral in \eqref{zgravdef_mt}, which is independently justified from the bulk perspective. }

\section{Details on bulk length overlap calculations}

\subsection{Average of length overlaps}
\label{app:length_details}

In this Appendix, we explicitly  resum the dominant contributions to the sum \eqref{genus_sum_mt} for $\ell \sim e^{S_0/3}$. The Weil-Petersson volume $V_{g, 1}(b)$ is given by \cite{Mirzakhani:2006eta} (see for instance Appendix~A of ~\cite{netta_free_energy} for a discussion from the physics perspective),
\be  \label{WP}
V_{g,1}(b) = \sum_{p=0}^{3g-2} \frac{(2\pi^2)^p}{2^{3g-2-p} (3g-2-p)! p!} b^{2(3g-2-p)} \underbrace{\le(\int_{\overline{M}_{g, 1}} \psi_1^{3g-2-p} \kappa^p\ri)}_{c_{g, p}}
\ee
The factor in the parentheses 
is known as an intersection number on the moduli space of curves \cite{Mirzakhani:2006eta}. For our purposes, its main feature is that it is a constant that depends on $g$ and $p$ but is independent of $b$. We will refer to this constant as $c_{g, p}$ in the remaining discussion. 
 Putting \eqref{WP} into \eqref{genus_sum}, we have 
\begin{align}
\overline{\braket{\ell|V^{\dagger}V|\ell'}} - \delta(\ell-\ell') 
&=\sum_{g=1}^{\infty} \sum_{p=0}^{3g-2} e^{-2gS_0} \frac{(2\pi^2)^p}{2^{3g-3-p} (3g-2-p)! p!} \,  c_{g, p} \, f(\ell, \ell', 6g-3 - 2p) \label{b_integral}
\end{align}
where we have defined  
\be \label{c3}
f(\ell, \ell', \alpha) \equiv \int_0^{\infty} db \, b^{\alpha}K_0\le(4\sqrt{e^{-\ell} + e^{-\ell'} + 2e^{-(\ell+\ell')/2} \cosh(b/2)}\ri)\, . 
\ee
Let us now analyze the above function. When $\ell$, $\ell'$ are large, we can neglect the first two terms inside the square root relative to the third for most of the range of $b$, so that we have 
\be \label{c4}
K_0\le(4 \sqrt{e^{-\ell}+ e^{-\ell'} + 2 e^{-(\ell+\ell')/2} \cosh(b/2)}\ri) \approx K_0(4 e^{-(\ell+\ell')/4 + b/4})
\ee
Then 
\be 
f (\ell, \ell', \alpha) \approx \int_0^{\infty} db K_0(4e^{-\bar \ell/2+b/4}) b^{\alpha}, \quad \bar \ell = \frac{\ell+\ell'}{2}\, . \label{f_simple}
\ee
 Now for a given $\bar\ell$, the integrand $K_0(4 e^{-\bar\ell/2+b/4}) b^{\alpha}$ is zero at $b=0$, and then rapidly grows for $b> 0$. It has a peak at $b^{\ast} \approx 2\bar\ell$, and becomes almost negligible for $b>2 \bar \ell$. There are two competing sources for the dominant contribution: the entire region from $b=0$ to $b\approx 2\bar\ell$, where the argument of $K_0$ is small, and the value around the saddle point $b^{\ast}$. In most of the region from $b=0$ to $b=2\bar\ell$, we can use the small argument expansion of $K_0(x)$: 
\be 
K_0(x) =  - \log (x/2) + O(1)
\ee
With this approximation, the contribution to \eqref{f_simple} is 
\be 
f(\ell, \ell',  \alpha) \approx \int_0^{2\bar\ell} (\bar\ell/2 - b/4 - \log 2) b^{\alpha} db  = \frac{(2\bar\ell)^{\alpha+2}}{4(\alpha+1)(\alpha+2)}(1 + O(1/\bar\ell)) \label{f_l_alpha}
\ee
The contribution from the saddle-point $b^{\ast}$ can be checked to be proportional to $(\bar \ell)^{\alpha + 3/2}$, which is subleading compared to \eqref{f_l_alpha}. The above analysis involves various approximations, but one can confirm the accuracy of \eqref{f_l_alpha} by checking numerically that the ratio between the full integral in \eqref{c3} and the approximation on the RHS of \eqref{f_l_alpha} approaches 1 for large $\ell, \ell'$ for any $\alpha$.

As we explain in Appendix~\ref{app:length_details}, when either $\ell$ or $\ell'$ is large, $f$ can be approximated as 
\be 
f(\ell, \ell', \alpha) \approx \frac{(2\bar \ell)^{\alpha+2}}{4(\alpha+1)(\alpha+2)}, \quad \bar\ell \equiv \frac{\ell + \ell'}{2} \, . \label{f_l_alpha_mt}
\ee
Putting \eqref{f_l_alpha_mt} back into \eqref{b_integral} with $\alpha = 6g-3-2p$, we have 
\begin{align}
\overline{\braket{\ell|V^{\dagger}V|\ell'}} - \delta(\ell-\ell')   = \sum_{g=1}^{\infty} L^{6g}\sum_{p=0}^{3g-2}  (2 \bar \ell)^{-2p-1}C_{g, p} ,  \label{39}
\end{align}
where 
\be 
L \equiv   {\bar \ell}e^{-S_0/3} \, , \quad C_{g, p}\equiv \, \, \frac{ 8^{g}2^{p+1} (2\pi^2)^p c_{g,p}}{p! (3g-2-p)! (6g-2-2p)(6g-1-2p)} \, . \label{c10} 
\ee
As discussed in the main text, when $\ell\sim e^{S_0/3}$, all but the $p=0$ terms in the above sum can be ignored due to the 
$({\bar\ell})^{-2p}$ factor in the summand.  In terms of the Weil-Petersson volume \eqref{WP}, this corresponds to only  keeping the term with the highest power of $b$: this approximation is also known as the ``Airy limit,'' and the corresponding  volumes are called ``Airy volumes''~\cite{witten, Saad:2022kfe}
\be \label{airy_volume_1}
c_{g, 0}= \frac{1}{24^g g!}, \quad \quad V_{g, 1}^{\rm Airy}(b) = \frac{b^{6g-4}}{2^{3g-2} 24^g (3g-2)! g! }\, . 
\ee
 Hence, in this limit, we get the following approximation to \eqref{genus_sum}: 
\begin{align}
\overline{\braket{\ell|V^{\dagger}V|\ell'}} - \delta(\ell-\ell') &= \frac{1}{\ell} \sum_{g=1}^{\infty} \frac{\le(\frac{L^6}{3}\ri)^{g} }{g! (3g-2)! (6g-2)(6g-1)} \, . 
\label{final_airysum}
\end{align}
The above sum can be performed analytically, and we find the following exact expression for the correction to the averaged length overlaps in the Airy limit: 
\be 
\overline{\braket{\ell|V^{\dagger}V|\ell'}} - \delta(\ell-\ell')  =\frac{1}{60\ell} L^6~ {}_1F_4\le(\frac{5}{6};\frac{4}{3}, \frac{5}{3}, \frac{11}{6}, 2; \frac{L^6}{81} \ri)  \, . \label{1F4_growth} 
\ee

The correction \eqref{1F4_growth} shows a rapid growth when the parameter $L$ is large. To see this, note that 
for a large value of its last argument, ${}_1F_4$ has the following behavior: 
\be 
{}_1 F_{4}(a; b_1, b_2, b_3, b_4; y) \approx \frac{\prod_i \Gamma(b_i)}{4 \sqrt{2} \pi^{3/2} \Gamma(a)} y^{\frac{a-\sum_i b_i -3}{4}} e^{4 y^{1/4}} (1 + O(y^{-1/4})), \quad y\gg 1\, .  \label{1F4}
\ee
Using \eqref{1F4}, 
we have for large $x$:
\begin{align}
\overline{\braket{\ell|V^{\dagger}V|\ell'}} - \delta(\ell-\ell') &= \frac{x^2}{60\bar \ell}{}_1 F_{4}\le(\frac{5}{4}; \frac{4}{3}, \frac{5}{3}, \frac{11}{6}, 2; \frac{L^6}{81}\ri)\nn  
& \approx \frac{x^{-1/4}}{8 \sqrt{2\pi} \bar \ell} e^{\frac{4}{3} L^{3/2}} \,  , \quad L \gg 1 \, \label{lhs} \\
& \sim {\rm exp}\le(\frac{4}{3} \le(\ell e^{-S_0/3}\ri)^{\frac{3}{2}}\ri), \quad L\gg 1 \, . \label{lhs2}
\end{align}

\subsection{Variance of length overlaps}
\label{app:variance_details}

In this Appendix, we  evaluate the sum in \eqref{var_sum} for the variance of length inner products, again in the Airy limit where $\ell, e^{S_0/3}$ are large and $\ell\sim e^{S_0/3}$. \eqref{var_sum} can be more explicitly written as 
\begin{align}
{\rm Var}(\ell, \ell') &=  4 e^{-2S_0} \int_0^{\infty} db_1 b_1 \int_0^{\infty} db_2 b_2  K_0(4 e^{-\bar\ell/2 + b_1/4}) K_0(4 e^{-\bar\ell/2 + b_2/4}) \frac{\delta(b_1-b_2)}{b_1}\nn 
&+ 4e^{-2S_0}\sum_{g=1}^{\infty} e^{-2gS_0}  \int_0^{\infty} db_1 b_1 \int_0^{\infty} db_2 b_2  K_0(4 e^{-\bar\ell/2 + b_1/4}) K_0(4 e^{-\bar\ell/2 + b_2/4}) V_{g, n=2}(b_1, b_2) \label{var_sum_eqn}
\end{align}
In the above expression, we have already used the approximation in \eqref{c4} under the assumption that $\ell$, $\ell'$ are large. 

It is straightforward evaluate the first term in \eqref{var_sum_eqn}, coming from the $g=0$ trumpet geometry in \eqref{var_sum}, by using the expansion of $K_0$ for small argument like in the discussion around \eqref{f_simple}-\eqref{f_l_alpha}. 
\begin{align} 
\text{genus zero term} &= 4e^{-2S_0} \int_0^{\infty} db \, b K_0(4 e^{-\bar\ell/2+b/4})^2 \approx   4 e^{-2S_0} \int_0^{2 \bar\ell} db \, b (\bar\ell/2-b/4)^2  \nn 
& = \frac{1}{3} e^{-2S_0} {\bar\ell}^4
\end{align}
In order to evaluate the remaining terms in the genus expansion \eqref{var_sum_eqn} in the Airy limit, we need to make use of the two-boundary Weil-Petersson volumes in the Airy limit: 
\begin{align}
V^{\rm Airy}_{g, n=2}(b_1, b_2) &= \frac{1}{2^{3g-1}}\sum_{\alpha=0}^{3g-1} \frac{1}{\alpha! (3g-1-\alpha))! } b_1^{2\alpha} b_2^{2(3g-1-\alpha)} \int_{\bar M_g, 2} \psi_1^{\alpha}  \psi_2^{3g-1-\alpha} \label{airy_full}\
\end{align}
This is the subset of terms in the full WP volume with the largest total power of $b_1$ and $b_2$. The topological invariants $\int_{\bar M_g, 2} \psi_1^{\alpha}  \psi_2^{3g-1 - \alpha}$ appearing in the above expression can be deduced from eqn. (70) of \cite{Eynard:2021zcj} to be 
\begin{align}
\int_{\bar M_g, 2} \psi_1^{\alpha} \psi_2^{3g-\alpha- 1} & = \frac{1}{24^g g!}\sum_{m=0}^{\text{min}(\alpha, 3g-1-\alpha)}  \binom{g}{m} \frac{(-3)^m}{2m+1}  \binom{3g-1-2m}{\alpha -m } \\
& = \frac{1}{g! 24^g} \binom{3g-1}{\alpha} \underbrace{{}_4F_{3}\le(\ha, -\alpha, 1+\alpha -3g, -g ;\frac{3}{2}, -\frac{(3g-1)}{2}, - \frac{(3g-2)}{2} ; \frac{3}{4}\ri)}_{F_{g, \alpha}}  \label{c20}
\end{align}
The ${}_4F_3$ factor in the above expression lies between 0 and 1 for all $\alpha$ in the allowed range $0 \leq \alpha \leq 3g-1$ for any $g$. In particular, this factor does not grow exponentially with $g$ and $\alpha$ and hence will not play an important role in our analysis below. We will abbreviate it as $F_{g, \alpha}$ in the remaining expressions to keep in mind that it has some weak dependence on $g$ and $\alpha$. 

Putting \eqref{airy_full} into \eqref{var_sum_eqn}, and evaluating the integrals over $b_1$ and $b_2$ using \eqref{f_l_alpha},  we find 
\be 
{\rm Var}(\ell, \ell')_{g\geq 1} = 4 e^{-2S_0}\sum_{g=1}^{\infty} \frac{e^{-2gS_0} (2\bar\ell)^{6g+4}}{16 \times g! 24^g \times 2^{3g-1}} \sum_{\alpha=0}^{3g-1} \frac{(3g-1)!}{\le(\alpha!(3g-1-\alpha)!\ri)^2} F_{g, \alpha} \frac{1}{(2\alpha +2) (2\alpha+3) (6g-2\alpha)(6g-2\alpha+1)} \label{c21}
\ee
Now taking $g$ to be large (which should be sufficient to capture the asymptotic large-$L$ behavior of the sum over all $g$) and using Stirling's approximation for the factorials in the sum over $\alpha$, we can see using the saddle-point approximation over $\alpha$ that the largest contribution is from $\alpha \approx 3g/2$, and the saddle-point value is 
\be 
\frac{(3g-1)!}{\le(\alpha!(3g-1-\alpha)!\ri)^2}\bigg|_{\alpha= 3g/2} \approx \frac{e^{3g \log 3g - 3g}}{e^{4 \times (\frac{3g}{2}\log \frac{3g}{2} - \frac{3g}{2})}} = \frac{2^{6g}}{e^{3g\log 3g- 3g}} \approx \frac{64^{g}}{(3g)!}  
\ee
Putting this back into \eqref{c21}, we find 
\be 
{\rm Var}(\ell, \ell')_{g\geq 1} = e^{-2S_0} {\bar\ell}^4 \sum_{g=1}^{\infty} \frac{\le(\frac{64L^6}{3}\ri)^{g}}{g!(3g!)} f(g) , \quad L\equiv  \bar\ell e^{-S_0/3}\, . 
\ee
where we have combined all factors with some power law dependence on $g$ into the function $f(g)$. 

The precise details of the function $f$ are beyond the scope of our analytical calculation. However, for  $f(g)\propto g^n$ for any integer power $n$, the genus sum is proportional to a hypergeometric function of the form ${}_p F_{p+3}(...;... ; \frac{64 L^6}{81})$. Other cases, where $f(g)$ has terms with fractional powers of $g$, can again be upper- and lower-bounded by ${}_p F_{p+3}(...;... ; \frac{64 L^6}{81})$. Using the asymptotic form of this hypergeometric function when its final argument is large, we can see that the leading behavior of $\rm Var(\ell, \ell')$ is \eqref{var_ell}. 

\subsection{Corrections to  closed universe inner products} \label{app:closeduniversevar}

Using the JT gravity path integral, the genus expansion for the inner product between two closed universe states in pure JT gravity with lengths $b_1$ and $b_2$ is given by 
\begin{align}
\overline{\lb b_1 |V^{\dagger}V| b_2\rb}-\frac{1}{b_1}\delta(b_1-b_2) = C + D 
\end{align}
where $C$ is the contribution from connected geometries, 
\begin{align}
C &= 
\raisebox{-.35cm}{\includegraphics[width=0.3\textwidth]{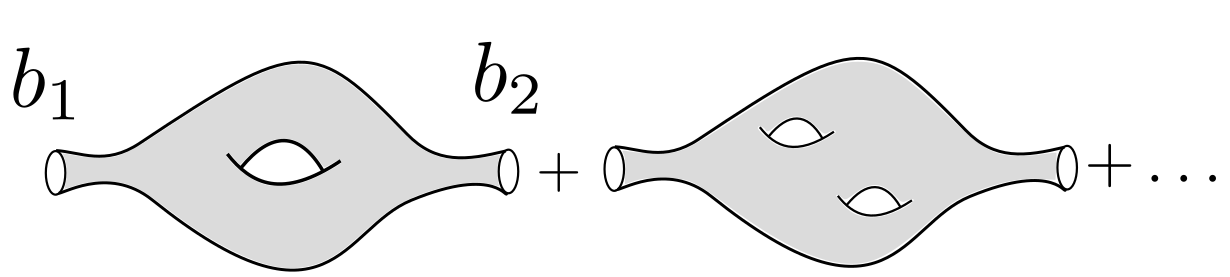}} \nn 
& = \sum_{g=1}^{\infty} e^{-2S_0g}  V_{g, 2}(b_1, b_2) \label{conn}
\end{align}
and $D$ is the contribution from disconnected geometries, 
\begin{align}
D & = \lr{\sum_{g=1}^{\infty} e^{-(2g-1)S_0}  V_{g, 1}(b_1)} \lr{\sum_{g=1}^{\infty} e^{-(2g-1)S_0}  V_{g, 1}(b_2)}  = \raisebox{-.5cm}{\includegraphics[width=0.3\textwidth]{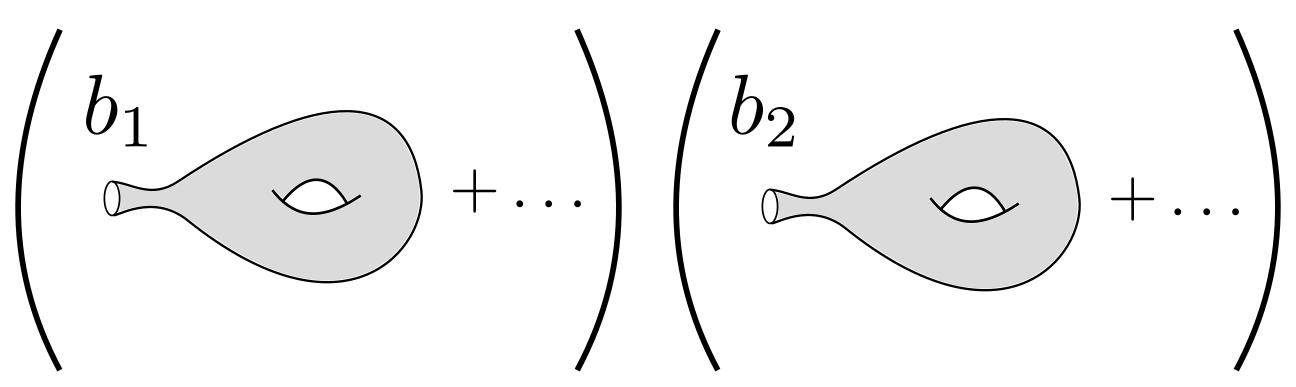}} \label{disc}
\end{align} 

Let us first analyze the disconnected contribution \eqref{disc}, using the Airy volume \eqref{airy_volume_1} assuming that $b_1, b_2 \sim e^{S_0/3}$. 
The sum over genus in each of the factors in \eqref{disc} then becomes convergent, and evaluates to the following closed-form expression: 
\be
\sum_{g=1}^{\infty} e^{-(2g-1)S_0}  V_{g, 1}^{\rm Airy}(b) = \frac{b^2 e^{S_0}}{48} {}_0 F_3(\frac{2}{3},\frac{4}{3}, 2; \frac{b^6 e^{-2 S_0}}{5184}) \sim \exp\le({\frac{(b e^{-S_0/3})^{3/2}}{6\sqrt{2}}}\ri), \quad (b e^{-S_0/3})\gg 1
\ee
so that for $b_1, b_2 \gtrsim e^{S_0/3}$, 
the leading contribution to $D$ grows as 
\be 
D \sim \exp\le({\frac{(b_1 e^{-S_0/3})^{3/2}}{6\sqrt{2}}}+{\frac{(b_2 e^{-S_0/3})^{3/2}}{6\sqrt{2}}}\ri), \label{c29}
\ee

Next, let us evaluate the connected contribution \eqref{conn}, again using the two-boundary Airy volumes  $V_{g, 2}^{\rm Airy}(b_1, b_2)$ in  \eqref{airy_full}-\eqref{c20}: 
\be 
C = \sum_{g=1}^{\infty} e^{-2S_0g} \frac{1}{2^{3g-1}} \frac{1}{g!24^g}\sum_{\alpha=0}^{3g-1} b_1^{2\alpha} b_2^{2(3g-1-\alpha)}  \frac{(3g-1)!} {\le(\alpha! (3g-1-\alpha))!\ri)^2 } F_{g, \alpha}
\ee
Now to find the asymptotic form of the summand for large $g$, we can again use Stirling's approximation together with the saddle-point approximation for the sum over $\alpha$ like in the discussion below \eqref{c21}. We now find that the saddle-point value is at $\alpha = 3gc$, where $c = \frac{b_1}{b_1+ b_2}$. Combining all factors, we find that for large $g$, 
\be 
\sum_{\alpha=0}^{3g-1}b_{1}^{2\alpha}  b_{2}^{2(3g-1-\alpha)} \frac{(3g-1)!}{\le(\alpha!(3g-1-\alpha)!\ri)^2} F_{g, \alpha}   \approx \frac{(2 \bar b )^{6g-2}}{(3g)!} F_{g, 2gc}, \quad \bar b \equiv \frac{b_1 + b_2}{2} \, . 
\ee
Putting this back into the sum over $g$, we have  
\begin{align}
\overline{\lb b_1 |V^{\dagger}V| b_2\rb}-\frac{1}{b_1}\delta(b_1-b_2)  
& = \frac{e^{-2S_0}}{{\bar b}^2} \sum_{g=1}^{\infty} \le( \frac{e^{-2S_0} (2\bar b)^6}{3}\ri)^g \frac{1}{g!(3g)!} h(g)
\end{align}
for some power law function $h(g)$. By the same reasoning as in the previous section, this expression can be approximated as a hypergeometric function of the form ${}_p F_{p+3}(...;...;\frac{(2 \bar b e^{-S_0/3})^6}{81}) \sim \exp\le(\frac{2\sqrt{2}}{3} (\bar b e^{-S_0/3})^{3/2}\ri)$. Comparing to the disconnected contribution in \eqref{c29}, we see that the connected contribution gives the dominant contribution to the closed universe inner product, which scales as 
\be 
\overline{\lb b_1 |V^{\dagger}V| b_2\rb}-\frac{1}{b_1}\delta(b_1-b_2) \sim \exp\le(\frac{2\sqrt{2}}{3} (\bar b e^{-S_0/3})^{3/2}\ri), \qq \bar b e^{-S_0/3} \gg 1 \label{finalb}
\ee

 The interpretation of the above result is that the closed universe wavefunction becomes very sensitive to negative energies when the universe is very big \cite{Usatyuk:2024isz}. 
In terms of the matrix integral, the above quantity can be interpreted as \cite{classifying_bc}~\footnote{More correctly, an infinite counterterm should be subtracted from the RHS to incorporate the fact that there are no contributions to the gravitational path integral from the disk topology with the closed geodesic boundary conditions.  See for example \cite{Blommaert:2021fob}. This counterterm does not affect our discussion. }
\be 
\overline{\lb b_1 |V^{\dagger} V| b_2\rb}  = \overline{\Tr[\cos(b \sqrt{H})]\Tr[\cos(b' \sqrt{H})]} 
\ee
We can see that the presence of negative energies in certain members of the ensemble gives a large variance to the quantity $\Tr \cos(b \sqrt{H})$, which becomes exponentially large in the presence of negative energies. This is the underlying reason for the exponential growth \eqref{finalb}.

\section{Variance of length overlaps from the Airy Kernel} 
\label{sec:variance_airy}

\subsection{Conventions for the Airy density of states and the Airy kernel} \label{airy_section}

Close to the edge of the spectrum at $E=0$, the normalized disk density of states in JT gravity (i.e., the leading order density of states in the $e^{-S_0}$ expansion) has the form 
\be 
\lim_{E \to 0}\rho_0(E) =  \lim_{E\to 0} \frac{1}{2\pi^2}\sinh(2\pi \sqrt{2E})  = \frac{\sqrt{2}}{\pi} \sqrt{E} \label{airy0}
\ee
In the case where the leading-order  normalized density of states is given by  
\be 
\tilde\rho_0(\lambda) = \frac{1}{\pi} \sqrt{\lambda} \, ,  \label{rho0_lambda}
\ee
it is known that in terms of the rescaled variable 
\be 
\xi = - e^{2S_0/3} \lambda
\ee
the density of states $\tilde \rho(\lambda)$ at all orders in $e^{-S_0}$, and in the regime where $\xi$ is $O(1)$, is given by 
\begin{align}
&\overline{\tilde \rho(\lambda)} =   \rho_{\rm Airy}(\lambda_1)\equiv e^{2S_0/3} [\Ai'(\xi)^2- \xi \Ai(\xi)^2] \label{airy_1pt}\\
& \overline{\tilde \rho(\lambda_1)\tilde\rho(\lambda_2)}_{\rm c} \equiv   \overline{\tilde \rho(\lambda_1)\tilde\rho(\lambda_2)} - \overline{\tilde \rho(\lambda)} \, \, \overline{\tilde\rho(\lambda_2)}= \delta(\lambda_1-\lambda_2) \rho_{\rm Airy}(\lambda) - e^{4S_0/3} K(\xi_1,\xi_2)^2 \label{airy_twopt}
\end{align}
where $K(\xi_1, \xi_2)$ is the Airy kernel 
\be 
K_{\rm Airy}(\xi_1, \xi_2) = \frac{{\rm Ai'}(\xi_1){\rm Ai}(\xi_2)- {\rm Ai'}(\xi_2){\rm Ai}(\xi_1)}{\xi_1 - \xi_2} \, . \label{kairy2}
\ee
Using the asymptotic form of the Airy function for large negative arguments, it can be checked that \eqref{rho0_lambda} is recovered in the $S_0 \to \infty$ limit: 
\be 
\lim_{S_0\to \infty} \overline{\tilde \rho(\lambda)} = \frac{1}{\pi}\sqrt{\lambda}\, . 
\ee
Now to get the nonperturbative $\overline{\rho(E)}$ and $\overline{\rho(E_1)\rho(E_2)}$ for $\rho(E)$ in \eqref{airy0}, we can define a change of variables $E = 2^{-1/3}\lambda$, which ensures that 
\be 
\rho_0(E) dE = \tilde \rho_0(\lambda) d\lambda
\ee
We then also have at finite $S_0$  
\be 
\overline{\rho(E)} dE = \overline{\tilde \rho(\lambda)}d\lambda, \quad  \overline{\rho(E_1)\rho(E_2)}_{\rm c} dE_1 dE_2 = \overline{\tilde \rho(\lambda_1)\tilde \rho(\lambda_2)}_{\rm c} d\lambda_1 d \lambda_2  \label{airy_twopoint}
\ee
 which can be used to find averages of functions involving $E$.

\subsection{Saddle-point calculation for the variance}

Let us now evaluate the energy integral for the variance ${\rm Var}(\ell, \ell')$ of $\Delta(\ell, \ell')$, using \eqref{cov_energy} and the connected two point function of the density of states from \eqref{airy_twopt}. It is useful to expand the Airy functions for large $\xi$, to get 
\begin{align}
&\rho_{\rm Airy}(\lambda) \approx e^{2S_0/3}\frac{1}{8\pi \xi} e^{-\frac{4}{3}\xi^{3/2}} , \quad \xi \to \infty \label{d10}\\ 
&K_{\rm Airy}(\xi_1, \xi_2) \approx -\frac{1}{4\pi} \frac{e^{-\frac{2}{3}(\xi_1^{3/2} + \xi_2^{3/2})}}{(\xi_1^{1/2} +\xi_2^{1/2})\xi_1^{1/4} \xi_2^{1/4}} , \quad \xi_1, \xi_2 \to \infty \label{d11}
\end{align}
We will leave out the prefactors that do not scale exponentially with some power of $\ell$ or $e^{S_0}$ in the calculation below. 
Putting this into \eqref{cov_energy}, with the notation 
\be 
\xi_i = - e^{2S_0/3}\lambda_i=: e^{2S_0/3} z_i, 
\ee
(i.e., positive $z_i$ corresponds to negative energy). We can again consider the contribution from the negative energy part of the integral, and expand the overlaps for large $\ell$, so that we have 
\begin{align} 
{\rm Var}(\ell, \ell') &= \int_{-\infty}^{\infty}d\lambda_1\int_{-\infty}^{\infty}d\lambda_2\overline{\rho(\lambda_1) \rho(\lambda_2)}_{\rm c} \braket{\ell|2^{-1/3}\lambda_1} \braket{2^{-1/3}\lambda_1|\ell'} \braket{\ell|2^{-1/3}\lambda_2} \braket{2^{-1/3}\lambda_2|\ell'} \\
& \supset \int_0^{\infty}dz_1 \int_0^{\infty}dz_2 \overline{\rho(\lambda_1) \rho(\lambda_2)}_{\rm c}\,  e^{2^{1/3}(\ell + \ell')(\sqrt{z_1}+ \sqrt{z_2})}
\end{align}
Now we 
can write $\text{Var}(\ell, \ell')$ as a sum of two terms, coming respectively from the two terms in \eqref{airy_twopoint}: 
\be 
\text{Var}(\ell, \ell') = A - B \label{138}
\ee
where 
\begin{align}
A &= \int_0^{\infty} dz {\rm exp}\le(-\frac{4}{3} \le( e^{2S_0/3}z\ri)^{3/2} + 2^{4/3}(\ell + \ell')\sqrt{z}\ri),\\
B &= \int_0^{\infty} dz_1 \int_0^{\infty} dz_2 {\rm exp}\le(-\frac{4}{3}\le(\le(e^{2S_0/3}z_1\ri)^{3/2}+ \le(e^{2S_0/3}z_2\ri)^{3/2}\ri)+ 2^{1/3}(\ell + \ell')(\sqrt{z_1} + \sqrt{z_2})\ri)\, . 
\end{align}
For $A$, the saddle-point value of $z$ is $z^{\ast}= 2^{1/3}\bar \ell e^{-S_0}$, from which we get 
\be 
A \sim {\rm exp}\le( \frac{8\sqrt{2}}{3} \le(\bar \ell e^{-S_0/3}\ri)^{3/2}\ri)\, . 
\ee
For $B$, the saddle-point values are at 
$z_1^{\ast}=z_2^{\ast}= 2^{-2/3}\bar \ell e^{-S_0}$, which gives 
\be 
B \sim {\rm exp}\le( \frac{8}{3} \le(\bar \ell e^{-S_0/3}\ri)^{3/2}\ri)\, . 
\ee
Hence, $A$ is the dominant contribution, and we obtain the scaling in \eqref{var_ell}.

\section{Length growth in TFD state with negative energy contributions}
\label{app:tfd_airy}

In this Appendix, we revisit the energy integral of \eqref{e1e2}, now extended to include negative energies. Like in the calculations in Sec.~\ref{sec:airy_mt} and Appendix~\ref{sec:variance_airy}, we will focus on the contribution to the integral from negative energies $E \sim e^{-2S_0/3}$, which is relevant for $\ell \gtrsim e^{S_0/3}$. We have 
\be
\lb \ell(t) \rb = \int_{-\infty}^\infty d \ell  \ell \overline{|\braket{\psi_{\beta + it}|V^{\dagger}V|\ell}|^2} \label{l_t_integral}
\ee
where
\be
\overline{|\braket{\psi_{\beta + it}|V^{\dagger}V|\ell}|^2} = \int_{-\infty}^{\infty} d\lambda_1 \int_{-\infty}^{\infty} d\lambda_2 \, e^{-\frac{1}{2^{4/3}}\le(2it(\lambda_1-\lambda_2) + \beta(\lambda_1 + \lambda_2)\ri)} \le(\overline{\rho(\lambda_1)}\, \, \overline{\rho(\lambda_2)} + \overline{\rho(\lambda_1)\rho(\lambda_2)}_{\rm c} \ri) \braket{2^{-1/3}\lambda_1|\ell}\braket{\ell|2^{-1/3}\lambda_2} \label{length_eqn}
\ee
Let us now use the averages of the density of states and its two-point function from \eqref{airy_1pt}-\eqref{airy_twopt}. Note that the first term of the connected two-point function \eqref{airy_twopt} is a contact term proportional to $\delta(\lambda_1-\lambda_2)$. Since the time-dependence in \eqref{length_eqn} is proportional to $\lambda_1-\lambda_2$, this term of the two-point function gives a time-independent contribution, which can be ignored in our analysis. A similar contribution was also ignored in the calculation of \cite{volume_luca}. 

Now, let us combine the contributions from the disconnected term and second term of \eqref{airy_twopt}, expand the Airy kernel for large positive $\xi$ as in \eqref{d10}-\eqref{d11}, and use the approximation \eqref{K_largel_mt} for the $\braket{\ell|E}$ factors at large $\ell$. Then the contribution to the above integral from negative energies, in terms of the variables $z_i = -\lambda_i$, is given by:  
\begin{align} 
\overline{|\braket{\psi_{\beta + it}|V^{\dagger}V|\ell}|^2}_{\rm negative~E}  & \sim \int_0^{\infty}dz_1 \int_0^{\infty}dz_2  \,(z_1 + z_2 - 2 \sqrt{z_1z_2}) \nn 
&\quad \quad \times \exp\lr{-\frac{4}{3}(z_1^{3/2}+z_2^{3/2})e^{S_0}+2^{1/3} \ell  (\sqrt{z_1}+\sqrt{z_2})+\frac{1}{2^{4/3}}(2 i t (z_1 - z_2)+\beta (z_1 + z_2)  } \label{e3}
\end{align}
In the above expression, we have ignored all prefactors that have power law scaling with $\ell$ and $e^{S_0}$, except the factor in the first line of \eqref{e3}, which comes from the combination of the connected and disconnected contributions to the density of states. We will see below that it is important to keep this prefactor in order to see that we get a non-zero result for the above quantity at all times, including early times. The remaining prefactors which we have ignored do not give competing contributions. 

Let us first solve the saddle-point equation for the second line of \eqref{e3} without including the prefactors. 
We get two solutions for each energy variable: 
\begin{align}
z_1 &=\frac{e^{-2S_0}}{2^{2/3} \times 32} \left(16 e^{S_0} \ell - 4 t^2 + 4 i t \beta + \beta^2 \pm \sqrt{(2 i t + \beta)^2 \left(32 e^{S_0} \ell - 4 t^2 + 4 i t \beta + \beta^2\right)}\right)\\
&=\frac{1}{2^{5/3}} e^{-S_0} \ell \pm e^{-3S_0/2} \sqrt{\ell} f(t,\beta) + \ldots \nonumber
\end{align}
where in the second line we have written the most important terms in the saddle for large $S_0$ and assuming $\ell \gtrsim e^{S_0/3}$. The saddle for the second energy variable is simply the complex conjugate of these saddles, $z_2 = (z_1)^*$. The saddles are within the regime of validity of the large $\xi$ expansion of the  Airy kernel for $\ell \gtrsim e^{S_0/3}$, and  are offset into the complex plane by an exponentially small amount $\epsilon \sim e^{-3S_0/2} \sqrt{\ell}$ since $f(t,\beta)t$ is a complex number for $t>0$.

To evaluate the integral, the contour for $E_1$ can be shifted up to $\mathbb{R}+i \epsilon$ to pick up the saddle above the real axis at $z_1=a+i \epsilon$. A steepest descent analysis reveals the remaining integral over $E_2$ picks up the conjugate saddle $z_2=a-i \epsilon$, which gives a real valued answer as required for a real length.\footnote{Similarly, if the contour was deformed downward to $\mathbb{R}-i \epsilon$ we would pick up the other set of saddles, which give the same result at leading order.} Evaluating the integral at the saddle, we find the following result at leading order: 
\be
\overline{|\braket{\psi_{\beta + it}|V^{\dagger}V|\ell}|^2}_{\rm negative~E}  \sim \exp \lr{\frac{2^{3/2}}{3} (e^{-S_0/3}\ell)^{3/2} } \label{e5}
\ee

Hence, the integrand of \eqref{l_t_integral} exponentially grows with length due to negative energies, starting at $\ell \sim e^{S_0/3}$, leading to a divergent integral.

In order to get a non-zero answer from the above analysis, it appears to be important that $t>0$, so that the difference between the  saddle-point energies  $z_1 -z_2 = 2 i \epsilon \neq 0$, and as a result the prefactor in the first line of \eqref{e3} is non-zero. However, a more careful  analysis of the prefactor reveals that we still get a non-zero answer at $t=0$, with the same leading behavior \eqref{e5}. 
The key point is that this prefactor repels the two saddle points from each other. Anticipating that $z_1$ and $z_2$ are close at leading order, we include the prefactor in the exponential. The most important terms in the exponent are then
\begin{align}
(\text{integrand}) \sim & \exp\lr{-\frac{4}{3}(z_1^{3/2}+z_2^{3/2})e^{S_0}+2^{1/3} \ell  (\sqrt{z_1}+\sqrt{z_2})+\log((\sqrt{z_1}-\sqrt{z_2})^2)}\nn 
 \sim  &\exp \lr{  -\frac{4}{3} (x_1^3 + x_2^3) + 2^{1/3} L (x_1 + x_2) + \log((x_1-x_2)^2)}
\end{align}
where we have further changed variables to the $O(1)$ quantities $x_i\equiv (z_i e^{2S_0/3})^{1/2}$ to simplify the discussion, and as before $L = \ell e^{-S_0/3}$. 
The saddle-point equations for $x_i$ do not have analytic solutions, but it can be easily seen that they scale as $x_i \sim 2^{-5/6} \sqrt{L}$, and the difference should be approximately bounded as $|x_1-x_2| \gtrsim 1/L$. Hence,  this more careful analysis  reproduces the exponential growth \eqref{e5} for all times including $t=0$.

\section{Doubly nonperturbative density of states and one-eigenvalue instantons}
\label{app:instanton}

Beyond the universal Airy regime $|E|\sim e^{-2S_0/3}$, the density of states in JT gravity is no longer given by \eqref{airy_1pt}. Recall that the saddle-point values of $E$ in calculations involving the Airy density of states in Sec.~\ref{sec:airy_mt} and Appendix~\ref{sec:variance_airy}  lie at $E^{\ast} = \ell e^{-S_0}$, so that these calculations are reliable only for $\ell \sim  e^{S_0/3}$. For larger $|E|$, the density of states at $E<0$ is known from non-perturbative instanton effects in the JT gravity matrix integral~\cite{sss}. The leading  contribution to the density of states in this region is given by 
\begin{align}
&\overline{\rho(E)} = \frac{1}{-8 \pi E} \exp (-V_{\mathrm{eff}}(E)), \qq E<0  \label{f1}\\
  &  V_{\mathrm{eff}}(E)=\frac{e^{S_0}}{4 \pi^3}[\sin (2 \pi \sqrt{-2E})-2 \pi \sqrt{-2E} \cos (2 \pi \sqrt{-2E})] \, . \label{f2}
\end{align}
We can check that the leading behavior of \eqref{f2} for small $|E|$ is $\approx \frac{4\sqrt{2}}{3}e^{S_0} (-E)^{3/2} = \frac{4}{3}\xi^{3/2}$, so that \eqref{f1} smoothly interpolates with the large $\xi$ expansion of the universal Airy density of states \eqref{d10}. For negative $E$ with larger magnitude, $V_{\rm eff}(E)$ shows oscillations and takes large positive as well as negative values. The large negative values of $V_{\rm eff}(E)$ indicate a nonperturbative instability in JT gravity, i.e., the leading-order density of states $\rho_0(E)$ does not provide a good approximation to the exact density of states \eqref{f1} if the integration contour for $E$ is taken to be the real line.

\begin{figure}[!h]
    \centering
    \includegraphics[width=0.35\linewidth]{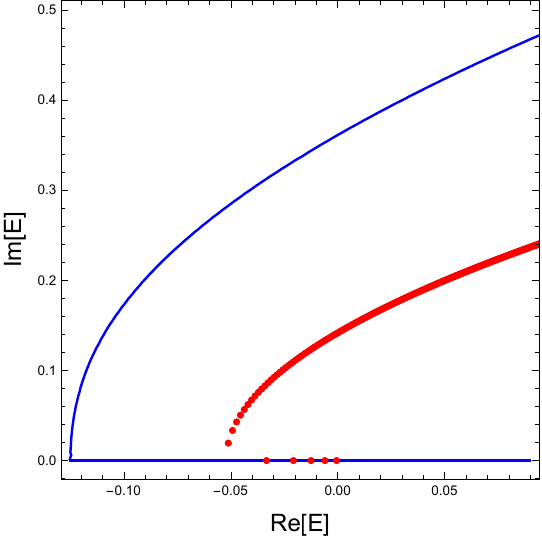}
    \caption{The complex energy contour $\Gamma$ for the JT gravity matrix integral is shown in blue. The contour is  obtained by  following the steepest descent line of $V_{\rm eff}(E)$, starting from its first maximum on the negative real axis at $E=-1/8$, into the upper half plane.  The saddle-points $E^{\ast}(k)$ of \eqref{f3} for various values of $k \equiv \bar \ell e^{-S_0}$ from $k=0.001$ to $k=3$ are shown in red. For small negative $k \lesssim 0.047 $, $E^{\ast}(k)$ lies at increasingly negative values on the real line as we increase $k$. For $k \gtrsim0.047$,  $E^{\ast}(k)$ moves into the upper half plane and its real part becomes less negative, and eventually positive,  on increasing $k$ further.}
    \label{fig:saddlepoint_k}
\end{figure}

\cite{sss} provided a prescription to avoid this instability by deforming the contour of the energy integral away from the real axis  after the first maximum of $V_{\rm eff}(E)$, to follow the steepest ascent contour of $V_{\rm eff}(E)$ into the upper half plane towards infinity. See the blue line in Fig.~\ref{fig:saddlepoint_k} for the resulting integration contour $\Gamma$. Alternatively, one can also follow the steepest ascent contour of $V_{\rm eff}(E)$ into the lower half plane to get the contour $\Gamma'$, which is the reflection of $\Gamma$ across the real line. For quantities that are expected to be real, it is standard to consider the sum of the integrals along $\Gamma$ and $\Gamma'$. 

With this deformed contour, the prescription for the full  contribution to the averaged overlap  $\overline{\braket{\ell|V^{\dagger}V|\ell'}}$ from negative and complex energies is then given by  (like in previous calculations, we ignore the prefactors in the integrand other than the exponential contributions)
\be 
\overline{\braket{\ell|V^{\dagger}V|\ell'}} = \ha \le[\int_{\Gamma}   dE +  \int_{\Gamma'}   dE\ri] \exp(\underbrace{-V_{\rm eff}(E) + \sqrt{2}(\ell + \ell') \sqrt{-E}}_{I(E, \ell, \ell')}) \label{f3}
\ee
The saddle point equation for \eqref{f3} is 
\be \label{f4}
\frac{1}{2\pi} \sqrt{-2E} \sin(2 \pi \sqrt{-2E}) = k, \qq k \equiv \bar\ell e^{-S_0} \, .  
\ee
For $k \ll 1$, the saddle point $E^{\ast}(k)$ is along the negative energy axis, as expected from the agreement of $\rho(E)$ with the Airy density of states in this limit. On increasing $k$, the saddle-point moves into the upper half plane. See Fig.~\ref{fig:saddlepoint_k}.

\begin{figure}[!h]
    \centering
    \includegraphics[width=0.45\linewidth]{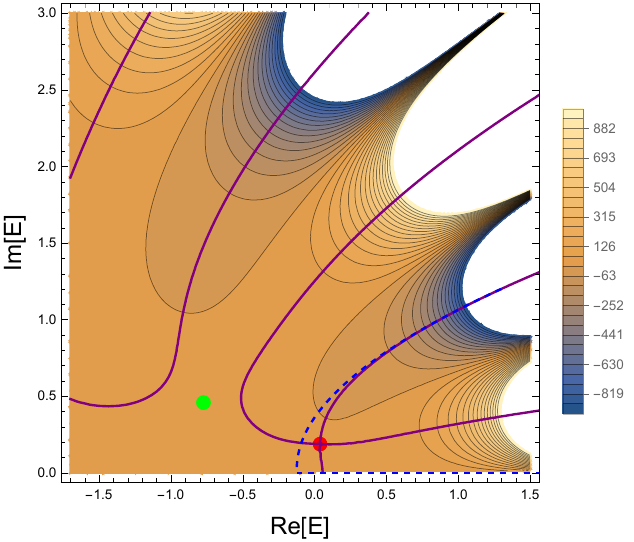} \includegraphics[width=0.45\linewidth]{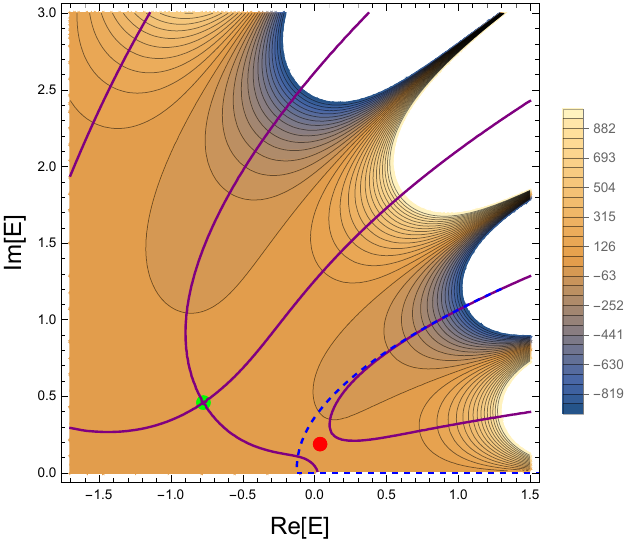}
    \caption{For $k=1$, we show the saddle-point value of $E$ in the complex plane  that gives the dominant contribution to the integral over $\Gamma$ (shown in the blue dashed contour) as the red point, and one example of another saddle-point that does not contribute to the integral in green. The purple lines are lines of constant $\Im f(E)$, where $f(E)\equiv -\frac{V_{\rm eff}(E)}{e^{S_0}} + 2k\sqrt{-2E}   $ is the exponent in \eqref{f3}. The contour plot shows the value of $\Re[f(E)]$ , which allows us to distinguish the steepest descent contours (along which $\Re[f(E)]$ decreases fastest) from the steepest ascent  contours (along which $\Re[f(E)]$ increases fastest) among the purple lines. The left and right plots respectively show these steepest ascent/descent contours for the red and green points. We see that for the red point, the steepest ascent contour intersects $\Gamma$ while the steepest descent contour asymptotes to $\Gamma$, indicating that it contributes to the integral. For the green saddle-point, the steepest ascent contour does not intersect $\Gamma$ and the steepest descent contour asymptotes in a direction far from $\Gamma$, indicating that it does not contribute.}
    \label{fig:saddlepoint_check}
\end{figure}

The integral in \eqref{f3} also has other complex saddle-points besides the ones shown in Fig.~\ref{fig:saddlepoint_k}. However, one can check using steepest ascent/ descent conditions that the saddle-points shown in Fig.~\ref{fig:saddlepoint_k} are the only ones that contribute to the integral along the contour $\Gamma$. We explain this reasoning in more detail for the example of $k=1$ in Fig.~\ref{fig:saddlepoint_check}.  

\begin{figure}[!h]
    \centering
\includegraphics[width=\linewidth]{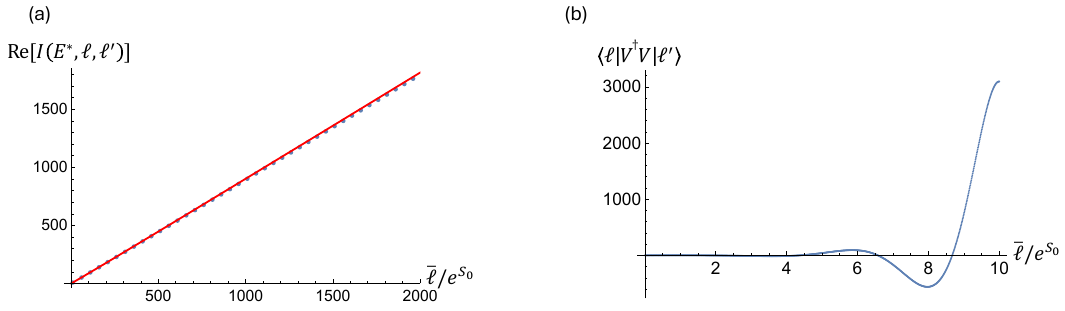}
    \caption{In (a), the blue data points show the real part of the exponent of the integrand in \eqref{f3} evaluated at the saddle-point value of $E$ (found numerically), and the red line shows the analytic estimate $\bar \ell(1- \frac{1}{\log(8\pi^2\bar \ell e^{-S_0})})$ for this quantity from \eqref{f6}.  (b) shows the numerical evaluation of the full quantity $\braket{\ell|V^{\dagger}V|\ell'}$ on including the imaginary part in the exponent and adding contributions from $\Gamma$ and $\Gamma'$. The resulting value of the overlap shows large oscillations as a function of $\bar \ell$ for large $\bar \ell$ (we show only up to $\ell e^{-S_0}=10$ as an illustration, as the numerical values rapidly become very large for larger $\ell$).}
    \label{fig:instanton_result}
\end{figure}

On plugging the saddle-point solutions shown in Fig.~\ref{fig:saddlepoint_k} into the exponent in \eqref{f3}, we find the behaviour of $\braket{\ell|V^{\dagger}V|\ell'}$ shown in Fig.~\ref{fig:instanton_result}. Note that for very small $k
\lesssim 10^{-2}$ (not shown in the plot), we have an initial behavior $\sim k^{3/2}$ of the log of the overlap, which smoothly interpolates with the Airy behavior~\eqref{deviation_result}.

We can understand this analytically as follows. From the saddlepoint equation \eqref{f4} we can redefine $\sqrt{-2 E}=z$ giving us $z\sin(2\pi z) = 2k$. If we are in the regime $k \gg 1$ and we assume, as the numerics suggest, that the imaginary part $\Im[z]< 0$, then the saddlepoint equation reduces to $2\pi i z e^{2\pi i z} = -8 \pi^2 k$. The solutions to this equation are labelled by an integer $n$ and are given by the Lambert function, $z=\frac{1}{2\pi i} W_n(-8\pi^2 k)$. The most dominant saddle in our case turns out to be $n=0$, and the asymptotics of the Lambert function are known and given by:\footnote{Note that we pick the value \eqref{f5} of the square root of $E^{\ast}(k)$, and not its negative, as the branch cut of the the square root is assumed to go from 0 to $-\infty$ throughout.} 
\be \label{f5}
\sqrt{-2 E^* (k)}\equiv z=\frac{1}{2\pi i} W_0(-8\pi^2 k)\approx \frac{1}{2}-\frac{1}{2 \log(8 \pi^2 k)}-\frac{i}{2\pi} \lr{ \log(8 \pi^2 k ) - \log \log(8 \pi^2 k) } + \ldots
\ee
One can solve for the energy to find that at large $k$ it is in the upper right quadrant. We find that $V_{\rm eff}(E^{\ast}(k))$ is subleading for large $k$ compared to the the second term $2 \bar \ell \sqrt{-2E(k^{\ast})}$ in  the exponent of \eqref{f4}. We therefore have, on adding the complex-conjugate contributions from $\Gamma$ and $\Gamma'$,
\be
\braket{\ell| V^{\dagger} V |\ell'} \approx \exp \lr{ e^{S_0} k\lr{1-\frac{1}{\log(8 \pi^2 k)}} - \frac{i}{\pi} e^{S_0} k \log(8 \pi^2 k) + \ldots }, \qq k \gg 1 \label{f6}
\ee

The overall magnitude of $\braket{\ell|V^{\dagger}V|\ell'}$ grows exponentially with $\ell$ (see Fig.~\ref{fig:instanton_result} (a)). The imaginary part leads to oscillations with $\ell$, so that we have a qualitative form 
\be
\braket{\ell|V^{\dagger}V|\ell'} \sim e^{\bar \ell} \cos\le(\frac{1}{\pi}\bar \ell \log \bar \ell\ri)
\ee
of the correction to the inner product at large $\ell$.


\bigskip
\bigskip

\bibliography{biblio.bib}

@ARTICLE{netta_free_energy,
       author = {{Engelhardt}, Netta and {Fischetti}, Sebastian and {Maloney}, Alexander},
        title = "{Free energy from replica wormholes}",
      journal = {\prd},
         year = 2021,
        month = feb,
       volume = {103},
       number = {4},
          eid = {046021},
        pages = {046021},
          doi = {10.1103/PhysRevD.103.046021},
archivePrefix = {arXiv},
       eprint = {2007.07444},
 primaryClass = {hep-th},
       adsurl = {https://ui.adsabs.harvard.edu/abs/2021PhRvD.103d6021E}
}

@ARTICLE{zhenbin,
       author = {{Yang}, Zhenbin},
        title = "{The quantum gravity dynamics of near extremal black holes}",
      journal = {Journal of High Energy Physics},
         year = 2019,
        month = may,
       volume = {2019},
       number = {5},
          eid = {205},
        pages = {205},
          doi = {10.1007/JHEP05(2019)205},
archivePrefix = {arXiv},
       eprint = {1809.08647},
 primaryClass = {hep-th},
       adsurl = {https://ui.adsabs.harvard.edu/abs/2019JHEP...05..205Y}
}

@ARTICLE{classifying_bc,
       author = {{Goel}, Akash and {Iliesiu}, Luca V. and {Kruthoff}, Jorrit and {Yang}, Zhenbin},
        title = "{Classifying boundary conditions in JT gravity: from energy-branes to {\ensuremath{\alpha}}-branes}",
      journal = {Journal of High Energy Physics},
         year = 2021,
        month = apr,
       volume = {2021},
       number = {4},
          eid = {69},
        pages = {69},
          doi = {10.1007/JHEP04(2021)069},
archivePrefix = {arXiv},
       eprint = {2010.12592},
 primaryClass = {hep-th},
       adsurl = {https://ui.adsabs.harvard.edu/abs/2021JHEP...04..069G}
}

@article{saad,
  author = {{Saad}, Phil},
  title = "{Late Time Correlation Functions, Baby Universes, and ETH in JT Gravity}",
  journal = {arXiv e-prints},
  year = 2019,
  month = oct,
  eid = {arXiv:1910.10311},
  pages = {arXiv:1910.10311},
  archivePrefix = {arXiv},
  eprint = {1910.10311},
  primaryClass = {hep-th},
  adsurl = {https://ui.adsabs.harvard.edu/abs/2019arXiv191010311S}
}

@ARTICLE{witten,
  author        = {{Witten}, Edward},
  title         = "{Two-dimensional gravity and intersection theory on moduli space}",
  journal       = {Communications in Mathematical Physics},
  year          = 1991,
  volume        = {141},
  pages         = {153-209},
  doi           = {10.1007/BF01217730},
  archivePrefix = {arXiv},
  eprint        = {hep-th/9204083},
  adsurl        = {https://ui.adsabs.harvard.edu/abs/1991CMaPh.141..153W}
}

@article{stanford_witten,
  author       = {{Stanford}, Douglas and {Witten}, Edward},
  title        = "{Fermionic localization of the Schwarzian theory}",
  journal      = {Journal of High Energy Physics},
  year         = 2017,
  month        = oct,
  volume       = {2017},
  number       = {10},
  pages        = {008},
  doi          = {10.1007/JHEP10(2017)008},
  archivePrefix= {arXiv},
  eprint       = {1703.04612},
  primaryClass = {hep-th},
  adsurl       = {https://ui.adsabs.harvard.edu/abs/2017JHEP...10..008S}
}

@ARTICLE{eth_jt,
       author = {{Jafferis}, Daniel Louis and {Kolchmeyer}, David K. and {Mukhametzhanov}, Baur and {Sonner}, Julian},
        title = "{JT gravity with matter, generalized ETH, and Random Matrices}",
      journal = {arXiv e-prints},
         year = 2022,
        month = sep,
          eid = {arXiv:2209.02131},
        pages = {arXiv:2209.02131},
          doi = {10.48550/arXiv.2209.02131},
archivePrefix = {arXiv},
       eprint = {2209.02131},
 primaryClass = {hep-th},
       adsurl = {https://ui.adsabs.harvard.edu/abs/2022arXiv220902131J}
}

@ARTICLE{magan,
       author = {{Balasubramanian}, Vijay and {Lawrence}, Albion and {Mag{\'a}n}, Javier M. and {Sasieta}, Martin},
        title = "{Microscopic Origin of the Entropy of Black Holes in General Relativity}",
      journal = {Physical Review X},
         year = 2024,
        month = feb,
       volume = {14},
       number = {1},
          eid = {011024},
        pages = {011024},
          doi = {10.1103/PhysRevX.14.011024},
archivePrefix = {arXiv},
       eprint = {2212.02447},
 primaryClass = {hep-th},
       adsurl = {https://ui.adsabs.harvard.edu/abs/2024PhRvX..14a1024B}
}

@article{marolfmaxfield,
  author = {{Marolf}, Donald and {Maxfield}, Henry},
  title = "{Transcending the ensemble: baby universes, spacetime wormholes, and the order and disorder of black hole information}",
  eprint = {2002.08950},
  archivePrefix = {arXiv},
  primaryClass = {hep-th},
  doi = {10.1007/JHEP08(2020)044},
  journal = {JHEP},
  volume = {08},
  pages = {044},
  year = {2020}
}

@ARTICLE{volume_luca,
       author = {{Iliesiu}, Luca V. and {Mezei}, M{\'a}rk and {S{\'a}rosi}, G{\'a}bor},
        title = "{The volume of the black hole interior at late times}",
      journal = {Journal of High Energy Physics},
         year = 2022,
        month = jul,
       volume = {2022},
       number = {7},
          eid = {73},
        pages = {73},
          doi = {10.1007/JHEP07(2022)073},
       adsurl = {https://ui.adsabs.harvard.edu/abs/2022JHEP...07..073I}
}

@ARTICLE{eynard_review,
  author        = {{Eynard}, B. and {Kimura}, T. and {Ribault}, S.},
  title         = "{Random matrices}",
  journal       = {arXiv:1510.04430 [math-ph]},
  year          = {2015},
  eprint        = {1510.04430},
  archivePrefix = {arXiv},
  primaryClass  = {math-ph},
  adsurl        = {https://ui.adsabs.harvard.edu/abs/2015arXiv151004430E/abstract}
}

@ARTICLE{hartman_replica,
       author = {{Almheiri}, Ahmed and {Hartman}, Thomas and {Maldacena}, Juan and {Shaghoulian}, Edgar and {Tajdini}, Amirhossein},
        title = "{Replica wormholes and the entropy of Hawking radiation}",
      journal = {Journal of High Energy Physics},
         year = 2020,
        month = may,
       volume = {2020},
       number = {5},
          eid = {13},
        pages = {13},
          doi = {10.1007/JHEP05(2020)013},
archivePrefix = {arXiv},
       eprint = {1911.12333},
 primaryClass = {hep-th},
       adsurl = {https://ui.adsabs.harvard.edu/abs/2020JHEP...05..013A}
}

@ARTICLE{pssy,
       author = {{Penington}, Geoff and {Shenker}, Stephen H. and {Stanford}, Douglas and {Yang}, Zhenbin},
        title = "{Replica wormholes and the black hole interior}",
      journal = {Journal of High Energy Physics},
         year = 2022,
        month = mar,
       volume = {2022},
       number = {3},
          eid = {205},
        pages = {205},
          doi = {10.1007/JHEP03(2022)205},
       adsurl = {https://ui.adsabs.harvard.edu/abs/2022JHEP...03..205P}
}

@article{Susskind:2014moa,
  author = {Susskind, Leonard},
  title = {Computational Complexity and Black Hole Horizons},
  journal = {Fortschritte der Physik},
  volume = {64},
  pages = {24-43},
  year = {2016},
  doi = {10.1002/prop.201500093},
  eprint = {1402.5674},
  archivePrefix = {arXiv},
  primaryClass = {hep-th}
}

@article{Susskind:2014yaa,
  author = {Susskind, Leonard},
  title = {Entanglement is not Enough},
  journal = {Fortschritte der Physik},
  volume = {64},
  pages = {49-71},
  year = {2016},
  doi = {10.1002/prop.201500095},
  eprint = {1411.0690},
  archivePrefix = {arXiv},
  primaryClass = {hep-th}
}

@article{Stanford:2014jda,
  author = {Stanford, Douglas and Susskind, Leonard},
  title = {Complexity and Shock Wave Geometries},
  journal = {Phys. Rev. D},
  volume = {90},
  number = {12},
  pages = {126007},
  year = {2014},
  doi = {10.1103/PhysRevD.90.126007},
  eprint = {1406.2678},
  archivePrefix = {arXiv},
  primaryClass = {hep-th}
}

@ARTICLE{penington,
       author = {{Penington}, Geoffrey},
        title = "{Entanglement wedge reconstruction and the information paradox}",
      journal = {Journal of High Energy Physics},
         year = 2020,
        month = sep,
       volume = {2020},
       number = {9},
          eid = {2},
        pages = {2},
          doi = {10.1007/JHEP09(2020)002},
archivePrefix = {arXiv},
       eprint = {1905.08255},
 primaryClass = {hep-th},
       adsurl = {https://ui.adsabs.harvard.edu/abs/2020JHEP...09..002P}
}

@ARTICLE{hawking2,
       author = {{Hawking}, S.~W.},
        title = "{Breakdown of predictability in gravitational collapse}",
      journal = {\prd},
         year = 1976,
        month = nov,
       volume = {14},
       number = {10},
        pages = {2460-2473},
          doi = {10.1103/PhysRevD.14.2460},
       adsurl = {https://ui.adsabs.harvard.edu/abs/1976PhRvD..14.2460H}
}

@ARTICLE{islands,
  author  = {{Almheiri}, Ahmed and {Mahajan}, Raghu and {Maldacena}, Juan and {Zhao}, Ying},
  title   = "{The Page curve of Hawking radiation from semiclassical geometry}",
  journal = {Journal of High Energy Physics},
  year    = {2020},
  month   = {Mar},
  pages   = {149},
  doi     = {10.1007/JHEP03(2020)149},
  eprint  = {1908.10996},
  archivePrefix = {arXiv},
  primaryClass  = {hep-th},
  adsurl  = {https://ui.adsabs.harvard.edu/abs/2020JHEP...03..149A/abstract}
}

@ARTICLE{hawking1,
       author = {{Hawking}, S.~W.},
        title = "{Particle Creation by Black Holes}",
      journal = {Advanced Series in Astrophysics and Cosmology},
         year = 1993,
        month = jun,
       volume = {8},
        pages = {85-106},
          doi = {10.1142/9789812384935_0005},
       adsurl = {https://ui.adsabs.harvard.edu/abs/1993AdSAC...8...85H}
}

@ARTICLE{aemm,
       author = {{Almheiri}, Ahmed and {Engelhardt}, Netta and {Marolf}, Donald and {Maxfield}, Henry},
        title = "{The entropy of bulk quantum fields and the entanglement wedge of an evaporating black hole}",
      journal = {Journal of High Energy Physics},
         year = 2019,
        month = dec,
       volume = {2019},
       number = {12},
          eid = {63},
        pages = {63},
          doi = {10.1007/JHEP12(2019)063},
archivePrefix = {arXiv},
       eprint = {1905.08762},
 primaryClass = {hep-th},
       adsurl = {https://ui.adsabs.harvard.edu/abs/2019JHEP...12..063A}
}

@ARTICLE{akers,
       author = {{Akers}, Chris and {Lucas}, Andrew and {Vikram}, Amit},
        title = "{On the reconstruction map in JT gravity}",
      journal = {arXiv e-prints},
         year = 2025,
        month = jun,
          eid = {arXiv:2506.18975},
        pages = {arXiv:2506.18975},
          doi = {10.48550/arXiv.2506.18975},
archivePrefix = {arXiv},
       eprint = {2506.18975},
 primaryClass = {hep-th},
       adsurl = {https://ui.adsabs.harvard.edu/abs/2025arXiv250618975A}
}

@ARTICLE{nonpert,
       author = {{Iliesiu}, Luca V. and {Levine}, Adam and {Lin}, Henry W. and {Maxfield}, Henry and {Mezei}, M{\'a}rk},
        title = "{On the non-perturbative bulk Hilbert space of JT gravity}",
      journal = {Journal of High Energy Physics},
         year = 2024,
        month = oct,
       volume = {2024},
       number = {10},
          eid = {220},
        pages = {220},
          doi = {10.1007/JHEP10(2024)220},
archivePrefix = {arXiv},
       eprint = {2403.08696},
 primaryClass = {hep-th},
       adsurl = {https://ui.adsabs.harvard.edu/abs/2024JHEP...10..220I}
}

@ARTICLE{harlow_jafferis,
       author = {{Harlow}, Daniel and {Jafferis}, Daniel},
        title = "{The factorization problem in Jackiw-Teitelboim gravity}",
      journal = {Journal of High Energy Physics},
         year = 2020,
        month = feb,
       volume = {2020},
       number = {2},
          eid = {177},
        pages = {177},
          doi = {10.1007/JHEP02(2020)177},
       adsurl = {https://ui.adsabs.harvard.edu/abs/2020JHEP...02..177H}
}

@ARTICLE{sss,
       author = {{Saad}, Phil and {Shenker}, Stephen H. and {Stanford}, Douglas},
        title = "{JT gravity as a matrix integral}",
      journal = {arXiv e-prints},
         year = 2019,
        month = mar,
          eid = {arXiv:1903.11115},
        pages = {arXiv:1903.11115},
          doi = {10.48550/arXiv.1903.11115},
archivePrefix = {arXiv},
       eprint = {1903.11115},
 primaryClass = {hep-th},
       adsurl = {https://ui.adsabs.harvard.edu/abs/2019arXiv190311115S}
}

@ARTICLE{maladacena_eternal,
       author = {{Maldacena}, Juan},
        title = "{Eternal black holes in anti-de Sitter}",
      journal = {Journal of High Energy Physics},
         year = 2003,
        month = apr,
       volume = {2003},
       number = {4},
          eid = {021},
        pages = {021},
          doi = {10.1088/1126-6708/2003/04/021},
archivePrefix = {arXiv},
       eprint = {hep-th/0106112},
 primaryClass = {hep-th},
       adsurl = {https://ui.adsabs.harvard.edu/abs/2003JHEP...04..021M}
}

@ARTICLE{non_isometric,
       author = {{Akers}, Chris and {Engelhardt}, Netta and {Harlow}, Daniel and {Penington}, Geoff and {Vardhan}, Shreya},
        title = "{The black hole interior from non-isometric codes and complexity}",
      journal = {Journal of High Energy Physics},
         year = 2024,
        month = jun,
       volume = {2024},
       number = {6},
          eid = {155},
        pages = {155},
          doi = {10.1007/JHEP06(2024)155},
archivePrefix = {arXiv},
       eprint = {2207.06536},
 primaryClass = {hep-th},
       adsurl = {https://ui.adsabs.harvard.edu/abs/2024JHEP...06..155A}
}

@ARTICLE{sergio,
       author = {{Hern{\'a}ndez-Cuenca}, Sergio},
        title = "{Entropy and spectrum of near-extremal black holes: semiclassical brane solutions to non-perturbative problems}",
      journal = {Journal of High Energy Physics},
         year = 2025,
        month = may,
       volume = {2025},
       number = {5},
          eid = {20},
        pages = {20},
          doi = {10.1007/JHEP05(2025)020},
archivePrefix = {arXiv},
       eprint = {2407.20321},
 primaryClass = {hep-th},
       adsurl = {https://ui.adsabs.harvard.edu/abs/2025JHEP...05..020H}
}

@ARTICLE{airy_tale,
       author = {{Antonini}, Stefano and {Iliesiu}, Luca V. and {Rath}, Pratik and {Tran}, Patrick},
        title = "{Black Hole Airy Tail}",
      journal = {\prl},
         year = 2025,
        month = nov,
       volume = {135},
       number = {19},
          eid = {191501},
        pages = {191501},
          doi = {10.1103/ft96-b212},
archivePrefix = {arXiv},
       eprint = {2507.10657},
 primaryClass = {hep-th},
       adsurl = {https://ui.adsabs.harvard.edu/abs/2025PhRvL.135s1501A}
}

@article{Eynard:2021zcj,
    author = "Eynard, Bertrand and Lewa{\'n}ski, Danilo",
    title = "{A natural basis for intersection numbers.}",
    eprint = "2108.00226",
    archivePrefix = "arXiv",
    primaryClass = "math.AG",
    doi = "10.13137/2464-8728/35487",
    journal = "Rend.  Ist.  Mat.  Univ.  Trieste",
    volume = "55",
    pages = "6",
    year = "2023"
}

@article{Mirzakhani:2006eta,
    author = "Mirzakhani, Maryam",
    title = "{Weil-Petersson volumes and intersection theory on the moduli space of curves}",
    doi = "10.1090/S0894-0347-06-00526-1",
    journal = "J. Am. Math. Soc.",
    volume = "20",
    number = "01",
    pages = "1--24",
    year = "2007"
}

@ARTICLE{maxfield_turiaci1,
       author = {{Ghosh}, Animik and {Maxfield}, Henry and {Turiaci}, Gustavo J.},
        title = "{A universal Schwarzian sector in two-dimensional conformal field theories}",
      journal = {Journal of High Energy Physics},
         year = 2020,
        month = may,
       volume = {2020},
       number = {5},
          eid = {104},
        pages = {104},
          doi = {10.1007/JHEP05(2020)104},
archivePrefix = {arXiv},
       eprint = {1912.07654},
 primaryClass = {hep-th},
       adsurl = {https://ui.adsabs.harvard.edu/abs/2020JHEP...05..104G}
}

@ARTICLE{johnson,
       author = {{Johnson}, Clifford V.},
        title = "{Non-Perturbative JT Gravity}",
      journal = {arXiv e-prints},
         year = 2019,
        month = dec,
          eid = {arXiv:1912.03637},
        pages = {arXiv:1912.03637},
          doi = {10.48550/arXiv.1912.03637},
archivePrefix = {arXiv},
       eprint = {1912.03637},
 primaryClass = {hep-th},
       adsurl = {https://ui.adsabs.harvard.edu/abs/2019arXiv191203637J}
}

@article{firewalls,
  author        = {Stanford, Douglas and Yang, Zhenbin},
  title         = {Firewalls from wormholes},
  year          = {2022},
  eprint        = {2208.01625},
  archivePrefix = {arXiv},
  primaryClass  = {hep-th},
  adsurl        = {https://ui.adsabs.harvard.edu/abs/2022arXiv220801625S}
}

@article{Saad:2022kfe,
    author = "Saad, Phil and Stanford, Douglas and Yang, Zhenbin and Yao, Shunyu",
    title = "{A convergent genus expansion for the plateau}",
    eprint = "2210.11565",
    archivePrefix = "arXiv",
    primaryClass = "hep-th",
    doi = "10.1007/JHEP09(2024)033",
    journal = "JHEP",
    volume = "09",
    pages = "033",
    year = "2024"
}

@article{Miyaji:2025yvm,
    author = "Miyaji, Masamichi and Ruan, Shan-Ming and Shibuya, Shono and Yano, Kazuyoshi",
    title = "{Non-perturbative overlaps in JT gravity: from spectral form factor to generating functions of complexity}",
    eprint = "2502.12266",
    archivePrefix = "arXiv",
    primaryClass = "hep-th",
    reportNumber = "YITP-25-28",
    doi = "10.1007/JHEP06(2025)251",
    journal = "JHEP",
    volume = "06",
    pages = "251",
    year = "2025"
}

@article{Miyaji:2024ity,
    author = "Miyaji, Masamichi",
    title = "{Non-perturbative discrete spectrum of interior length and timeshift in two-sided black hole}",
    eprint = "2410.20662",
    archivePrefix = "arXiv",
    primaryClass = "hep-th",
    reportNumber = "YITP-24-139",
    doi = "10.1007/JHEP04(2025)190",
    journal = "JHEP",
    volume = "04",
    pages = "190",
    year = "2025"
}

@article{Mirzakhani:2006fta,
    author = "Mirzakhani, Maryam",
    title = "{Simple geodesics and Weil-Petersson volumes of moduli spaces of bordered Riemann surfaces}",
    doi = "10.1007/s00222-006-0013-2",
    journal = "Invent. Math.",
    volume = "167",
    number = "1",
    pages = "179--222",
    year = "2006"
}

@article{DeWitt:1967yk,
    author = "DeWitt, Bryce S.",
    editor = "Fang, Li-Zhi and Ruffini, R.",
    title = "{Quantum Theory of Gravity. 1. The Canonical Theory}",
    doi = "10.1103/PhysRev.160.1113",
    journal = "Phys. Rev.",
    volume = "160",
    pages = "1113--1148",
    year = "1967"
}

@book{mehta2004random,
  title={Random matrices},
  author={Mehta, Madan Lal},
  volume={142},
  year={2004},
  publisher={Elsevier}
}

@article{Usatyuk:2024isz,
    author = "Usatyuk, Mykhaylo and Zhao, Ying",
    title = "{Closed universes, factorization, and ensemble averaging}",
    eprint = "2403.13047",
    archivePrefix = "arXiv",
    primaryClass = "hep-th",
    doi = "10.1007/JHEP02(2025)052",
    journal = "JHEP",
    volume = "02",
    pages = "052",
    year = "2025"
}

@article{Hernandez-Cuenca:2024xlg,
    author = "Hern{\'a}ndez-Cuenca, Sergio and Valdes-Meller, Nico and Weng, Wayne Wei-en",
    title = "{A single geometry from an all-genus expansion in quantum gravity}",
    eprint = "2412.08799",
    archivePrefix = "arXiv",
    primaryClass = "hep-th",
    reportNumber = "MIT-CTP/5814",
    month = "12",
    year = "2024"
}

@article{McNamara:2020uza,
    author = "McNamara, Jacob and Vafa, Cumrun",
    title = "{Baby Universes, Holography, and the Swampland}",
    eprint = "2004.06738",
    archivePrefix = "arXiv",
    primaryClass = "hep-th",
    month = "4",
    year = "2020"
}

@article{Usatyuk:2024mzs,
    author = "Usatyuk, Mykhaylo and Wang, Zi-Yue and Zhao, Ying",
    title = "{Closed universes in two dimensional gravity}",
    eprint = "2402.00098",
    archivePrefix = "arXiv",
    primaryClass = "hep-th",
    doi = "10.21468/SciPostPhys.17.2.051",
    journal = "SciPost Phys.",
    volume = "17",
    number = "2",
    pages = "051",
    year = "2024"
}

@article{Harlow:2025pvj,
    author = "Harlow, Daniel and Usatyuk, Mykhaylo and Zhao, Ying",
    title = "{Quantum mechanics and observers for gravity in a closed universe}",
    eprint = "2501.02359",
    archivePrefix = "arXiv",
    primaryClass = "hep-th",
    reportNumber = "MIT-CTP/5824",
    doi = "10.1007/JHEP02(2026)108",
    journal = "JHEP",
    volume = "02",
    pages = "108",
    year = "2026"
}

@article{complexity_action,
  author = {{Brown}, Adam R. and {Roberts}, Daniel A. and {Susskind}, Leonard and {Swingle}, Brian and {Zhao}, Ying},
  title = "{Complexity, action, and black holes}",
  journal = {Physical Review D},
  year = 2016,
  volume = {93},
  number = {8},
  pages = {086006},
  doi = {10.1103/PhysRevD.93.086006},
  archivePrefix = {arXiv},
  eprint = {1512.04993},
  primaryClass = {hep-th},
  adsurl = {https://ui.adsabs.harvard.edu/abs/2016PhRvD..93h6006B}
}

@article{TracyWidom1994Airy,
  author  = {Tracy, Craig A. and Widom, Harold},
  title   = {Level-Spacing Distributions and the Airy Kernel},
  journal = {Communications in Mathematical Physics},
  volume  = {159},
  number  = {1},
  pages   = {151--174},
  year    = {1994},
  doi     = {10.1007/BF02100489}
}

@article{Abdalla:2025gzn,
    author = "Abdalla, Ahmed I. and Antonini, Stefano and Iliesiu, Luca V. and Levine, Adam",
    title = "{The gravitational path integral from an observer{\textquoteright}s point of view}",
    eprint = "2501.02632",
    archivePrefix = "arXiv",
    primaryClass = "hep-th",
    doi = "10.1007/JHEP05(2025)059",
    journal = "JHEP",
    volume = "05",
    pages = "059",
    year = "2025"
}

@article{Blommaert:2021fob,
    author = "Blommaert, Andreas and Iliesiu, Luca V. and Kruthoff, Jorrit",
    title = "{Gravity factorized}",
    eprint = "2111.07863",
    archivePrefix = "arXiv",
    primaryClass = "hep-th",
    doi = "10.1007/JHEP09(2022)080",
    journal = "JHEP",
    volume = "09",
    pages = "080",
    year = "2022"
}

\newpage

\end{document}